\documentclass[11pt]{article}
\pdfoutput=1

\usepackage{euscript}
\usepackage{amsfonts}
\usepackage{amsbsy}
\usepackage{epsfig}
\usepackage{amsthm}
\usepackage{amscd}
\usepackage{amstext}
\usepackage{verbatim}
\usepackage{cancel}
\usepackage{capt-of}
\usepackage{empheq}

\usepackage[nosort]{cite}
\usepackage{bm}
\usepackage{authblk}
\usepackage{esint}
\usepackage[T1]{fontenc}
\usepackage{mathdots}
\usepackage[centertableaux]{ytableau}

\usepackage[utf8]{inputenc}
\usepackage{graphicx}
\usepackage{physics}
\usepackage{mathtools}
\usepackage{bbold}

\usepackage{setspace}
\usepackage{colortbl}
\usepackage{xcolor}

\usepackage{amsmath}
\makeatletter
\renewcommand*\env@matrix[1][\arraystretch]{%
  \edef\arraystretch{#1}%
  \hskip -\arraycolsep
  \let\@ifnextchar\new@ifnextchar
  \array{*\c@MaxMatrixCols c}}
\makeatother
\usepackage{amssymb}
\usepackage{mathrsfs}
\usepackage{mathtools}
\usepackage{physics}
\usepackage{booktabs}
\usepackage{bbm}
\usepackage{float}

\usepackage{pgfplots}
\pgfplotsset{compat=1.15}
\usepackage{tikz}
\usetikzlibrary{external}
\usetikzlibrary{arrows}

\usepackage{subcaption}
\usepackage{graphicx}
\usepackage{mathtools}
\usepackage[many]{tcolorbox}
\tcbset{shield externalize}
\usepackage{xcolor, colortbl}
\usepackage[a4paper]{geometry}
\usepackage[hidelinks,linktocpage]{hyperref}
\hypersetup{
    colorlinks=true,
    linkcolor=blue,
    citecolor=blue,
    }

\usepackage{mathrsfs}

\def\ben{\begin{equation}}
\def\een{\end{equation}}
\def\half{{\textstyle{\frac{1}{2}}}}
\let\a=\alpha

\let\pa=\partial
\def\be{\begin{equation}}
\def\ee{\end{equation}}
\def\beq{\begin{equation}}
\def\eeq{\end{equation}}
\def\ba{\begin{array}}
\def\ea{\end{array}}

\def\dalemb#1#2{{\vbox{\hrule height .#2pt
       \hbox{\vrule width.#2pt height#1pt \kern#1pt
               \vrule width.#2pt}
       \hrule height.#2pt}}}

\newcommand{\bea}{\begin{eqnarray}}
\newcommand{\eea}{\end{eqnarray}}

\def\vep{{\varepsilon}}

\makeatletter
\newcommand*\bigcdot{\mathpalette\bigcdot@{.5}}
\newcommand*\bigcdot@[2]{\mathbin{\vcenter{\hbox{\scalebox{#2}{$\m@th#1\bullet$}}}}}
\makeatother

\def\R{{{\mathbb R}}}
\def\P{{{\mathbb P}}}

\def\Z{{{\mathbb Z}}}

\def\N{{{\mathbb N}}}
\def\Lag{{\mathcal{L}}}

\def\ocal{{\mathcal{O}}}

\title{The Conformal Primon Gas at the End of Time}
\author{Sean~A.~Hartnoll and Ming~Yang}
\affil{\it Department of Applied Mathematics and Theoretical Physics, \\
\it University of Cambridge, Cambridge CB3 0WA, UK
}
\date{}

\begin{document}

\maketitle

\begin{abstract}
The Belinksy-Khalatnikov-Lifshitz dynamics of gravity close to a spacelike singularity can be mapped, at each point in space separately, onto the motion of a particle bouncing within half the fundamental domain of the modular group. We show that the semiclassical quantisation of this motion is a conformal quantum mechanics where the states are constrained to be modular invariant. Each such state defines an odd automorphic $L$-function. In particular, in a basis of dilatation eigenstates the wavefunction is proportional to the $L$-function along the critical axis and hence vanishes at the nontrivial zeros.
We show that the $L$-function along the positive real axis is equal to the partition function of a gas of non-interacting charged oscillators labeled by prime numbers.
This generalises Julia's notion of a primon gas. Each state therefore has a corresponding, dual, primon gas with a distinct nontrivial set of chemical potentials that ensure modular invariance. We extract universal features of these theories by averaging the logarithm of the partition function over the chemical potentials. The averaging produces the Witten index of a fermionic primon gas.

\end{abstract}

\newpage

\tableofcontents

\newpage

\section{Introduction}

At a spacetime singularity the effective geometrical description of gravity breaks down and the microscopic degrees of freedom of our universe are expected to be revealed.
Dramatic singularities demarcating the `end of time' are likely present at the big bang and in the interior of black holes. Over 50 years ago, a remarkable paper by Belinksy, Khalatnikov and Lifshitz (BKL) argued that the classical dynamics of gravity would greatly simplify as such singularities are approached, even as the validity of the classical description breaks down \cite{Belinsky:1970ew}. It is tempting to ask whether this simplification is a harbinger of the emerging microscopic description 
\cite{Damour:2002et,Kleinschmidt:2009hv, Kleinschmidt:2009cv,belinski_henneaux_2017,Kleinschmidt:2022qwl}.

\begin{figure}[h]
    \centering
    \includegraphics[width=0.35\linewidth]{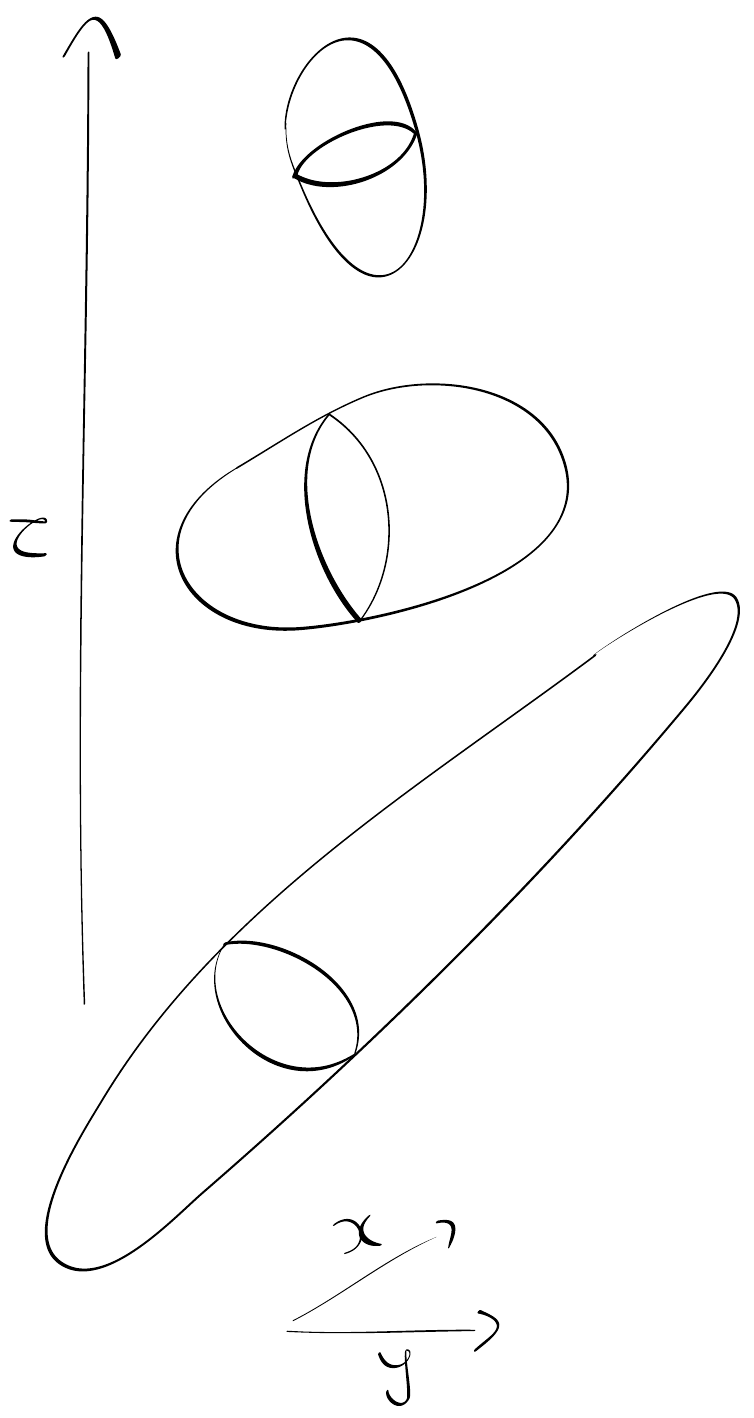}
    \hspace{1cm}
    \includegraphics[width=0.35\linewidth]{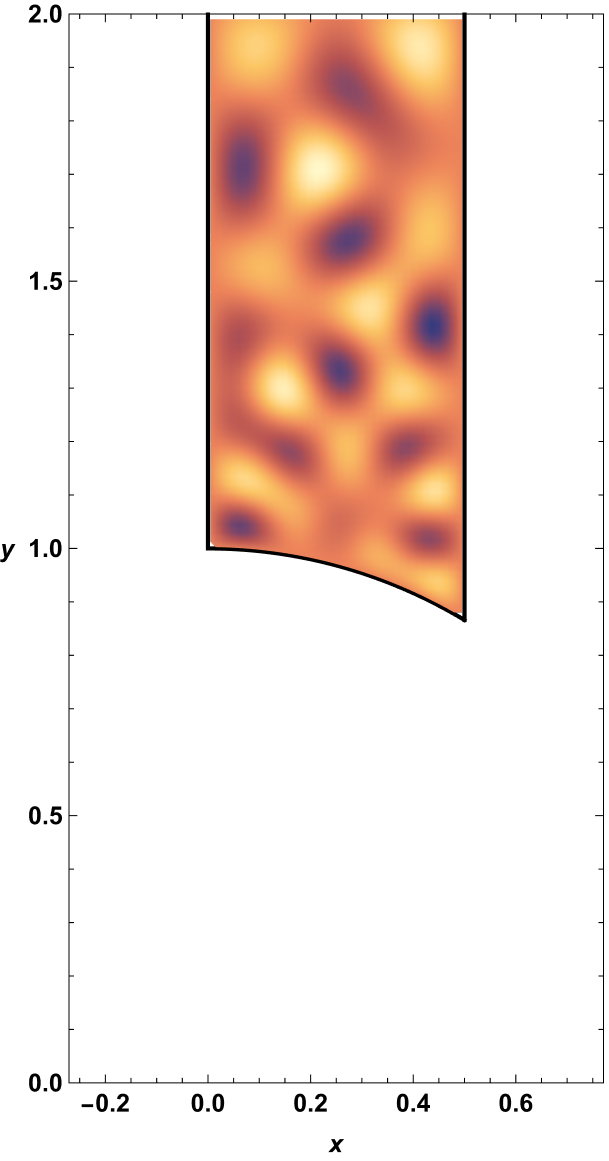}
    \caption{Left: schematic evolution of a local volume element in the BKL regime. The shape of the element, parametrised by $(x,y)$, evolves chaotically as the volume goes to zero at the singularity. The volume collapse is parameterised by $\tau$. Right: the semiclassical eigenfunctions of $\pa_\tau$ are odd automorphic forms of $SL(2,\Z)$. The eigenfunction shown has $\vep \approx 44.78$, plotted with data from \url{https://www.lmfdb.org/}. See \S\ref{sec:auto} for a more precise description.}
    \label{fig:shape}
\end{figure}

In the 90's it was understood that the mathematical framework underpinning the BKL dynamics of four dimensional Einstein gravity is arithmetic quantum chaos \cite{GRAHAM19951103}. This is because the BKL evolution of the spatial metric at each point in space, independently, can be mapped onto the dynamics of hyperbolic billiards in half of the fundamental domain of $SL(2,\Z)$ \cite{belinski_henneaux_2017}. This particular billiard problem is characterised by the existence of conserved Hecke operators that impart a number-theoretic character to the phase space \cite{Bogomolny:1992cj, sarnak1993arithmetic}.
In particular, the semiclassical wavefunctions of these billiard systems 
are automorphic forms that connect directly to deep mathematical conjectures including the Sato-Tate conjecture and Riemann hypothesis for automorphic $L$-functions --- see e.g.~\cite{Sarnak1987,n5, rudnick} and below. The step from cosmology to automorphic forms is illustrated in Fig.~\ref{fig:shape}.

Our first observation will be that BKL waveforms of the kind illustrated in Fig.~\ref{fig:shape} can be interpreted as states in a conformal quantum mechanics (CQM). This description is in the spirit of an AdS$_2$/CQM correspondence, see e.g.~\cite{Chamon:2011xk}, but with a few wrinkles. The Euclidean signature AdS$_2$ bulk is just the hyperbolic plane discussed above. This hyperbolic plane is not part of the BKL spacetime itself but rather a spatial slice of the superspace in which (the metric at each point of) the BKL spacetime evolves. The BKL superspace has an emergent time-translation invariance generated by $\pa_\tau$, emphasised in \cite{DeClerck:2023fax}.
There is a discrete set of eigenstates of this superspace time evolution operator, labeled by `energies' $\vep$. These states have a complex conformal weight $\Delta = \frac{1}{2} + i \vep$ and hence transform in a principal series representation of the conformal algebra. We go on to show 
that these states exhibits nontrivial mathematical and physical properties.

The key mathematical object that underpins both the bulk automorphic waveforms and the boundary CQM state is an automorphic $L$-function. The function $L(s)$ shares many properties with the Riemann zeta function, discussed in \S\ref{sec:autoL} and illustrated in Fig.~\ref{fig:L} below. For $\text{Re}(s) > 1$ the $L$-function can be represented as an Euler product over prime numbers. Upon analytic continuation to smaller $s$, the $L$-function is conjectured to have nontrivial zeros along a line $s = \frac{1}{2} + i t$. We will show in \S\ref{sec:conf} that if the CQM state is written in a basis of dilatation eigenstates $\ket{t}$, its wavefunction $\phi(t)$ is essentially the $L$-function along the critical line $\phi(t) \propto L(\frac{1}{2} + i t)$. This fact has some similarities to ideas proposed by Connes, Berry-Keating and Okazaki for the zeros of the Riemann zeta function, as we now explain.

The zeros of the Riemann zeta function have been associated to the eigenvalues of self-adjoint operators since comments made by Hilbert and P\'olya over a hundred years ago.
This connection has been reinforced by strong analogies between number theoretic and quantum mechanical phenomena including the Selberg trace formula, Montgomery's pair correlation conjecture and Berry's analysis of semi-classical quantum chaos. These are reviewed in \cite{berry1999riemann}. Three observations will be especially relevant for us. Firstly, as pointed out by Connes \cite{Connes1999}, a minus sign in the asymptotic formula for the density of zeros suggests that the zeros should be interpreted as an absorption spectrum. Secondly, as noted by Berry and Keating \cite{Berry1999}, the dilatation operator $xp$ has the same asymptotic spectral density as the Riemann zeros. Thirdly, as noted by Okazaki in this context \cite{Okazaki:2015lpa}, the natural setting for a dilatation operator is a conformal quantum mechanics. We have shown that the CQM state corresponding to an automorphic form vanishes, in a basis of dilatation eigenstates, at the nontrivial zeros of an $L$-function.

Our second observation is that the function $L(s)$, now along the real axis rather than the critical line, can also be extracted from the CQM state. This is done by taking overlaps of the state with a certain non-orthogonal collection of states $\ket{\psi_s}$. The Euler product representation implies that $L(s)$ can be interpreted as the partition function at inverse temperature $s$ of a non-interacting collection of bosonic oscillators, each labeled by a prime number $p$. For the Riemann zeta function this partition function is known as the primon gas \cite{julia1990statistical}. In \S\ref{sec:ds} we generalise the primon gas construction to automorphic $L$-functions. The new ingredient is that the oscillators are charged and each have an imaginary chemical potential. The chemical potentials associate a phase $\theta_p$ to each oscillator. These phases are fixed for a given eigenstate (i.e.~a given $\vep$), and are subtly correlated due to $SL(2,\Z)$ invariance. The Sato-Tate conjecture states that, on average, the phases are distributed with a $\sin^2\theta$ density for $\theta \in [0,\pi]$. We call this system the conformal primon gas as it can be thought of as a partition function dual to the CQM state. A state/partition function correspondence is familiar from gapped states with topological order \cite{RevModPhys.89.025005} or dS/CFT-type dualities \cite{Strominger:2001pn, Maldacena:2002vr}. In an AdS/CFT context a boundary partition function can also define a state in a black hole interior \cite{Hartnoll:2022snh}, or possibly more generally through $T^2$ deformation \cite{Araujo-Regado:2022gvw}. In a similar vein, it is tempting to think of the conformal primon gas as a partition function living on the black hole singularity.

To get a handle on the conformal primon gas partition function we average the logarithm of the partition function over the parameter $\vep$. This average is possible because, for each $p$, the phase $\theta_p$ follows a Kesten-McKay distribution over the set of conformal primon gases \cite{Sarnak1987}. The procedure here is somewhat analogous to averaging over CFT$_2$ data \cite{Collier:2019weq, Belin:2020hea, Chandra:2022bqq} which, like the $\{\theta_p\}$, is constrained by modular invariance. Remarkably, the averaged logarithm of the partition function is equal to the Witten index of a fermionic primon gas. The averaged partition function can be used to exhibit a divergence as $s \to \frac{1}{2}$, controlled by the asymptotic distribution of prime numbers. The divergence is due to the zero of odd automorphic $L$-functions at $s=\frac{1}{2}$. Averaging over an ensemble of partition functions also smooths out various strongly fluctuating sums that arise in so-called explicit formulae for $L$-functions, as we show in \S\ref{sec:explicit}.

A central theme throughout is that the data specifying an $L$-function can be encoded in either the collection of zeros $\{t_n\}$ or the collection of phases $\{\theta_p\}$ (equivalently, the corresponding Dirichlet/Fourier coefficients $c_p \equiv 2 \cos \theta_p$). As elaborated in \S\ref{sec:semi}, these distinct presentations of the function correspond to expressing the CQM state in different bases. Understanding the holographic dual of the singularity amounts to understanding how this data, that organises the semiclassical quantisation of gravity in the near-singularity regime, arises from a more complete quantum state incorporating microscopic degrees of freedom such as strings or matrices. This would be analogous to how the boundary gravitational Hamiltonian is uplifted to that of ${\mathcal N} = 4$ SYM theory in the AdS/CFT correspondence \cite{Maldacena:1997re}. The singularity is deep in the interior of spacetime, and not easily accessible to a boundary. The gravitational Hamiltonian is therefore defined relationally, in essence evolving the local shape of a spatial slice as a function of its volume. Our hope is that the emergent automorphic and conformal symmetries of the gravitational states may guide the search for a complete holographic description of the end of time.

\section{Automorphic waveforms from BKL dynamics}
\label{sec:auto}

\subsection{The Hamiltonian constraint for BKL dynamics}

In this section we will review how automorphic waveforms emerge as Wheeler-DeWitt wavefunctions for BKL dynamics. In general relativity, the local Hamiltonian vanishes. BKL argued that the form of this `Hamiltonian constraint' simplifies close to spacelike singularities in two ways \cite{Belinsky:1970ew}. Firstly, the dynamics of each spatial point decouples. Secondly, the scale factors for the spatial metric follow a free motion punctuated by sudden bounces. This dynamics, for each point in space separately, can be mapped onto a hyperbolic billiard problem, as we now describe following the notation in \cite{Damour:2002et, belinski_henneaux_2017}. We will simply list the pertinent facts.\footnote{
The regime of validity of BKL-like decoupling of spatial points within the fully nonlinear and inhomogeneous evolution of Einstein's equation is yet to be established. For some positive results that incorporate a degree of inhomogeneity, see e.g.~\cite{Garfinkle:2020lhb} for numerics and \cite{fournodavlos2023stable, fournodavlos2023asymptotically, Li:2024qjq} for rigorous results.}

The three-metric at a given point on a spatial slice is written in the Iwasawa decomposition
\be\label{eq:decom}
ds^2_3 = e^{-2\beta_1} \theta_1^2 + e^{-2\beta_2} \theta_2^2 + e^{-2\beta_3} \theta_3^2 \,,
\ee
where $\theta_1 = dx_1 + n_1 dx_2 + n_2 dx_3$, $\theta_2 = dx_2 + n_3 dx_3$ and $\theta_3 = dx_3$. The spatial coordinates are $\{x_1,x_2,x_3\}$. Close to a singularity, it is found that the $n_a$ freeze to constants and the nontrivial dynamics is given by the $\beta_a$. Let $\pi_a$ be the momentum conjugate to $\beta_a$, at a given point in space. To leading order close to a spacelike singularity, the Hamiltonian constraint is found to be
\be\label{eq:Ham1}
\left(\pi_1^2 + \pi_2^2 + \pi_3^2 \right) - \frac{1}{2} \left(\pi_1 + \pi_2 + \pi_3 \right)^2 + \Theta(\beta_1 - \beta_2) + \Theta(\beta_2 - \beta_3) + \Theta(- \beta_1) = 0 \,.
\ee
Here $\Theta(x)$ is an infinite step function that vanishes for $x < 0$ and is infinite for $x>0$. These `walls', that bound the configuration space, arise as the small volume (large $\sum_a \beta_a$) limit of exponential potentials. Note that while different points in space decouple, spatial gradients are important to obtain all of the walls.

We now perform the change of variables, with $y>0$,
\be
\beta_1 = \frac{e^\tau}{\sqrt{2}} \frac{x}{y} \,, \qquad \beta_2 = \frac{e^\tau}{\sqrt{2}} \frac{1-x}{y} \,, \qquad \beta_3 = \frac{e^\tau}{\sqrt{2}} \frac{x(x-1)+y^2}{y}\,.
\ee
This transformation maps the Minkowski superspace metric in (\ref{eq:Ham1}) to the Milne model metric.
In terms of $\{\tau,x,y\}$ and their conjugate momenta, the Hamiltonian constraint (\ref{eq:Ham1}) becomes
\be\label{eq:billiard}
\pi_\tau^2 = y^2 \left(\pi_x^2 + \pi_y^2 \right)  \qquad \text{with} \qquad 0 \leq x \leq \frac{1}{2} \quad \text{and} \quad x^2 + y^2 \geq 1 \,.
\ee
Here we see that $\pi_\tau^2$ is the Hamiltonian for a particle moving on the upper half plane representation of hyperbolic space. The $SL(2,\R)$ symmetry of the hyperbolic space is the $SO(1,2)$ Minkowski symmetry in (\ref{eq:Ham1}). In (\ref{eq:billiard}) we see that the particle is constrained to lie within precisely half of the fundamental domain of $SL(2,\Z)$, shown in Fig.~\ref{fig:shape} above. In the hyperbolic description (\ref{eq:billiard}) the superspace time $\tau$ generated by $\pi_\tau$ does not appear explicitly. This is only true in the near-singularity limit, where the walls are infinitely steep, and entails the conservation of $\pi_\tau$. As $\tau \to \infty$ the local volume element (\ref{eq:decom}) collapses, doubly exponentially, towards the singularity.

In a Wheeler-DeWitt (WDW) quantisation of the Hamiltonian constraint, the momenta are promoted to differential operators acting on a wavefunction $\Psi(\tau,x,y)$ in the canonical way \cite{DeWitt:1967yk}. 
There is an ordering ambiguity in this process, the most natural ordering is one where the Hamiltonian constraint becomes a Laplace operator with respect to the inverse DeWitt metric (see e.g.~\cite{DeClerck:2023fax} for further discussion). The time-independence of (\ref{eq:billiard}) means that one can decompose the solution into modes labelled by `energies' $\vep_k$  \cite{Graham:1990jd}. With the appropriate ordering, these obey
\be\label{eq:modes}
- y^2 \left(\frac{\pa^2}{\pa x^2} + \frac{\pa^2}{\pa y^2} \right) \Psi_k(x,y)  = \left(\frac{1}{4} + \vep_k^2 \right) \Psi_k(x,y) \,,
\ee
and must vanish at $x=0$, $x= \frac{1}{2}$ and on $x^2 + y^2 = 1$. The full wavefunction is found to be
\be
\Psi(\tau,x,y) = \sum_k \a_k \Psi_k(x,y) e^{-\tau/2}e^{i \vep_k \tau} \,.
\ee
Here the $\a_k$ are free coefficients.
These solutions have conserved and positive DeWitt norm --- see \cite{DeClerck:2023fax} for further discussion of this point. We will furthermore endow them with a representation-theoretic norm in \S\ref{sec:conf}. The remainder of our work is concerned with the solutions to (\ref{eq:modes}).

\subsection{Automorphic waveforms}

Equation (\ref{eq:modes}) lands us squarely in the field of arithmetic quantum chaos \cite{Bogomolny:1992cj, sarnak1993arithmetic}. For an overview of early work on WDW quantisation of cosmological billiards and arithmetic chaos, see \cite{GRAHAM19951103}. That discussion is mostly focused on the mixmaster minisuperspace \cite{Misner:1969hg}, which leads to a larger domain in the upper half plane.
The smaller domain that we will be interested in corresponds to a certain subsector of the minisuperspace case, and hence many results can be carried over directly. We can emphasise here that we are concerned with the full inhomogeneous BKL dynamics, there is no minisuperspace approximation. The modes (\ref{eq:modes}) exist independently at each point of the spatial slice. Separating variables, the general solution to (\ref{eq:modes}) is the Maa{\ss} cusp form
\be\label{eq:fourier}
\Psi_k(x,y) = \sum_{n=1}^\infty c^k_n \sqrt{y} K_{i \vep_k}(2 \pi n y) \sin (2 \pi n x) \,.   
\ee
Here $K$ is a modified Bessel function. The solutions (\ref{eq:fourier}) vanish as required on $x=0,\frac{1}{2}$. Imposing vanishing on $x^2+y^2=1$ determines the discrete spectrum of $\vep_k$ and the corresponding coefficients $c^k_n$ can be computed numerically \cite{n1,n2,n3,n4,n5}. Fig.~\ref{fig:shape} shows an illustrative waveform. More intricate examples can be found in \cite{em/1048610117}. While (\ref{eq:modes}) also admits a continuum spectrum, given by the Eisenstein series \cite{sarnak1993arithmetic}, this does not obey the Dirichlet boundary condition at $x=0$.

The set of functions obtained after imposing Dirichlet boundary conditions, as described above, are in correspondence with the odd automorphic forms of $SL(2,\Z)$. Firstly, vanishing at $x=0$ allows the waveforms to be extended to the full fundamental domain via
\be\label{eq:odd}
\Psi(-x,y) = - \Psi(x,y) \,.
\ee
Secondly, we can use transformations $\gamma \in SL(2,\Z)$ to extend the waveform from the fundamental domain to the full upper half plane as
\be\label{eq:auto1}
\Psi(\gamma(x,y)) = \Psi(x,y) \,.
\ee
Vanishing at $x=0$ together with automorphicity can be seen to imply vanishing on the other boundaries of the half-fundamental domain. Thus every solution to the Dirichlet problem defines an odd automorphic form and vice-versa. It will often be useful for us to consider the `unfolded' waveforms, extended in this way to the whole upper half plane.

The automorphic forms are eigenfunctions of Hecke operators \cite{Bogomolny:1992cj, sarnak1993arithmetic}. 
We will say more about Hecke operators in \S\ref{sec:sl2}. The fact that the Hecke operators commute with $SL(2,\Z)$ implies the Hecke relations, in which the coefficients $c_n^k$ are expressed in terms of the prime coefficients $c_p^k$, $p \in {\mathbb P}$. In a normalisation where $c_1^k=1$, the coefficients can be simplified as follows:
\begin{align}
c_{mn}^k & = c^k_m c^k_n && \text{when} \quad (n,m)=1 \,, \label{eq:hecke1} \\
c^k_{p^{n+1}} & =c^k_p c^k_{p^n} - c^k_{p^{n-1}} \,, &&  p \in {\mathbb P} \,. \label{eq:hecke2}
\end{align}
The first relation reduces the coefficient to a product of prime power coefficients, and the second relation further reduces these to prime coefficients. For example, $c^k_{18} = c^k_{9} c^k_{2} = ((c^k_{3})^2-1) c^k_{2}$.

The second relation (\ref{eq:hecke2}) can be solved explicitly. It is convenient to parameterise
\be
c^k_p \equiv 2 \cos \theta^k_p \,,
\ee
where we assume the Ramanujan conjecture stating that $0 \leq \theta_p^k \leq \pi$ is real, so that $|c_p^k| \leq 2$ \cite{Sarnak1987}, and then
\be\label{eq:recsol}
c^k_{p^n} = \frac{\sin ([n+1]\theta^k_p)}{\sin \theta^k_p} \,.
\ee
In this way, the semiclassical wavefunctions of a BKL universe are parameterised by a collection of angles $\{\theta_p^k\}$, labeled by prime numbers.

The angles $\theta_p^k$ obey highly nontrivial constraints due to the automorphicity condition (\ref{eq:auto1}). These angles are believed to be essentially random and, according to the Sato-Tate conjecture, distributed with a $\sin^2\theta$ density for $\theta \in [0,\pi]$ for a given $k$ \cite{Sarnak1987, n5}. This corresponds to a Wigner semi-circle distribution for the $c_p^k$. Given that hyperbolic billiards are classically ergodic systems, it is not entirely surprising that the Fourier coefficients in the wavefunction are random (cf.~\cite{MVBerry_1977}). The Hecke relations for arithmetic domains such as our one, however, imply that it is only the prime coefficients that are random.

To pre-empt possible confusions with similar functions arising in other contexts, let us emphasise two facts. Firstly, the waveforms are automorphic but are not holomorphic functions of $x + i y$. Secondly, the $c_p^k$ are not (generically) integer or rational numbers.

\section{Automorphic $L$-functions}
\label{sec:autoL}

A key fact for the remainder of our discussion is that the Hecke recursion relations allow the $c^k_n$ coefficients to be assembled, as a Dirichlet series, into an $L$-function. There is a separate $L$-function for each energy level $\vep_k$, defined with the corresponding coefficients. The Dirichlet series converges absolutely for $\text{Re}(s) > 1$ and, in that range, admits an Euler product representation
\begin{align}\label{eq:L}
L_k(s) \equiv \sum_{n=1}^\infty \frac{c^k_n}{n^s} & = \prod_{p \in \P}\sum_{m=0}^\infty \frac{c^k_{p^m}}{p^{ms}} \\
& = \prod_{p \in \P} \frac{1}{1-c^k_pp^{-s}+p^{-2s}} \equiv \prod_{p \in \P} L^{(p)}_k(s) \,. \label{eq:local}
\end{align}
The first step uses (\ref{eq:hecke1}) and the second step uses (\ref{eq:recsol}). The final expression defines the local $L$-functions $L^{(p)}_k(s)$. We may note the `chirality' factorisation
\be\label{eq:chiral}
L_k(s) = L_{k+}(s) L_{k-}(s) \,, \qquad L_{k\pm}(s) \equiv \prod_{p \in \P} \frac{1}{1-e^{\pm  i \theta^k_p}p^{-s}} \,.
\ee
For real $s$, $L_{k+(s)} = L_{k-}(s)^*$.

We will now enumerate some remarkable properties of the $L$-functions defined in (\ref{eq:L}). We restrict attention to $L$-functions associated to the odd parity waveforms in (\ref{eq:fourier}), as these are the ones that arise from BKL dynamics.
It is useful to introduce the xi function defined as \cite{Hafner1987, Epstein1985}
\be\label{eq:xi}
\xi_k(s) \equiv \frac{1}{2} \frac{1}{\pi^s} \Gamma\left(\frac{s+1+i\vep_k}{2} \right) \Gamma\left(\frac{s+1-i\vep_k}{2} \right) L_k(s) \,.
\ee
The xi function can be shown to have the reflection symmetry
\be\label{eq:inv}
\xi_k(s) = - \xi_k(1-s) \,,
\ee
which allows analytic continuation to the complex plane.
The xi function is entire and, analogously to the Riemann hypothesis, is believed only to have zeros along the critical line at
\be
s^k_n = \half + i t^k_n \,,
\ee
with $t^k_n \in \R$. The poles in the gamma functions in (\ref{eq:xi}) cancel out `trivial' zeroes in the $L$-function.
In particular this means that
\be
\xi_k(\half + i t) = \xi_k'(\half) \, t \prod_{n=1}^\infty \left(1 - \frac{t^2}{t_n^{k\,2}} \right) \,. \label{eq:zeros}
\ee
The zeros $\{t^k_n\}$, for a fixed $k$, share statistical properties with the eigenvalues of a random matrix \cite{RS, rudnick}. The reflection symmetry (\ref{eq:inv}) implies that $\xi_k(\half + i t)$ is pure imaginary.

The full waveform (\ref{eq:fourier}) can be recovered from the xi function via a Mellin transform \cite{Hafner1987, Epstein1985}
\be\label{eq:auto}
\Psi_k(\rho,\theta) = \frac{-1}{2\pi}\int_{-\infty}^\infty \xi_k \left(\half + i t \right) e^{i t \log \rho} K_k(\theta,t) dt \,, 
\ee
where $x + i y = \rho e^{i\theta}$, in the upper half plane so that $0 \leq \rho$ and $0 \leq \theta \leq \pi$, 
and
\be\label{eq:k}
K_k(\theta,t) = \frac{\cot \theta}{(\sin \theta)^{i t}}
\, {}_2F_1\left(\frac{3}{4} + \frac{i (t+\vep_k)}{2}, \frac{3}{4} + \frac{i (t-\vep_k)}{2}, \frac{3}{2}; - \cot^2 \theta \right) \,.
\ee
The function $K_k(\theta,t)$ obeys
the Schr\"odinger equation
\be
- \frac{d^2K_k}{d\theta^2} - \frac{1 + 4 \vep_k^2}{4 \sin^2\theta} K_k = - t^2 K_k \,,
\ee
and is odd about $\theta = \frac{\pi}{2}$. Taking an inverse Fourier transform of (\ref{eq:auto}) gives
\be\label{eq:inverse}
\xi_k\left(\half + i t \right) K_k(\theta,t) = - \int_0^\infty \Psi_k(\rho,\theta) \rho^{-1 - i t} d \rho \,.
\ee
In (\ref{eq:inverse}) the `unfolded' $\Psi_k(\rho,\theta)$, extending over the full upper half plane, is used. Using (\ref{eq:inverse}), modular invariance $\Psi_k(\rho,\theta) = \Psi_k(1/\rho,\pi - \theta)$, together with $K_k(\pi-\theta,-t) = - K_k(\theta,t)$, is seen to imply the reflection symmetry (\ref{eq:inv}) of the xi function.

\begin{figure}[h]
    \centering
    \includegraphics[width=0.5\linewidth]{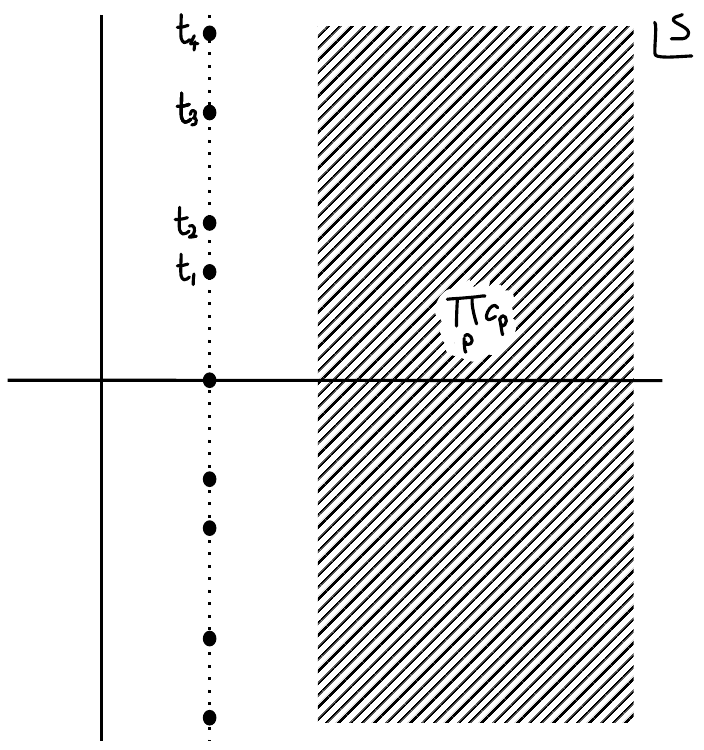}
    \caption{Schematic illustration of an automorphic $L$-function in the complex $s$ plane. The nontrivial zeros are shown as black dots along the line $s = \frac{1}{2} + i t$. The Euler product representation is established in the shaded region of absolute convergence of the Dirichlet series. Explicit formulae are obtained by considering contour integrals of $\xi'(s)/\xi(s)$, that pick up the nontrivial zeros, and then deforming the contour into the Euler product region --- see \S\ref{sec:explicit}.}
    \label{fig:L}
\end{figure}

We have seen how the $L$-function, and hence the automorphic waveform, can be specified either in terms of the collection of nontrivial zeros $\{t^k_n\}$ in (\ref{eq:zeros}) or in terms of the angles $\{\theta^k_p\}$ in (\ref{eq:chiral}). We have noted that both of these collections of numbers are (conjecturally) randomly distributed on average, and yet must obey subtle correlations. The existence of these two distinct product representations of the $L$-function
leads to `explicit formulae' relating the zeros and the primes \cite{rudnick}. We will elaborate on such formulae in \S\ref{sec:explicit}, meanwhile Fig.~\ref{fig:L} illustrates
the starting point.
In the remainder of this paper we will develop a physical approach to automorphic $L$-functions that encompasses these two different representations.

In the following \S\ref{sec:conf} we will use the $SL(2,\R)$ symmetry of the hyperbolic plane to interpret the unfolded waveform as a scattering state in a conformal quantum mechanics (CQM). We will understand this state to live on the boundary of the hyperbolic plane, in the spirit of an AdS$_2$/CQM correspondence \cite{Chamon:2011xk}. The spectrum $\{t_n\}$ of zeros are seen to be zeros for the dilatation operator in this CQM, reminiscent of ideas proposed by Connes \cite{Connes1999}, Berry-Keating \cite{Berry1999} and Okazaki \cite{Okazaki:2015lpa} for the Riemann zeros. In \S\ref{sec:semi} we show that a relationship between the spectrum of zeros and the Fourier coefficients, the `approximate functional equation' (or Riemann-Siegel formula for the case of the Riemann zeta function), follows from relating the wavefunction of the state in the dilatation and translation bases. In \S\ref{sec:ds} we return to the Euler product form of the $L$-function. This expression is naturally interpreted as the partition function of a dual `primon gas', generalising the observations of \cite{julia1990statistical} for the Riemann zeta function. The $\{\theta_p\}$ angles appear as chemical potentials of the dual partition function.

\section{Conformal quantum mechanics}
\label{sec:conf}

\subsection{$SL(2,\R)$ properties of the waveform}
\label{sec:sl2}

In this subsection we recast the properties of the waveform in terms of the $SL(2,\R)$ symmetry of the hyperbolic plane. This will then allow an interpretation of the waveform as a state in a conformal quantum mechanics. Most immediately, the Laplace equation (\ref{eq:modes}) on the hyperbolic upper half plane can be re-written in terms of the Casimir of $SL(2,\R)$. Let
\be\label{eq:gen1}
H = - i \pa_x \,, \qquad D = i (x \pa_x + y \pa_y) \,, \qquad K = - i\left((x^2 - y^2) \pa_x + 2 x y \pa_y\right) \,,
\ee
which obey the $SL(2,\R)$ algebra
\be
[D,H] = - i H \,, \qquad [D,K] = i K \,, \qquad [H,K] = 2 i D \,.
\ee
The $SL(2,\R)$ Casimir is then
\be
C_2 \equiv KH - D^2 + i D  = y^2 \left(\pa_x^2 + \pa_y^2 \right) \,,
\ee
and the Laplace equation (\ref{eq:modes}) becomes
\be\label{eq:cas}
C_2 \Psi_k = - \left(\frac{1}{4} + \vep_k^2 \right) \Psi_k = \Delta_k (\Delta_k - 1) \Psi_k \,.
\ee
Here the conformal weight
\be\label{eq:dep}
\Delta_k = \frac{1}{2} + i \vep_k \,,
\ee
indicating that the waveform is within a principal series representation of $SL(2,\R)$. This is not the more common representation considered in an AdS/CFT context, as it corresponds to bulk fields with mass squared below the Breitenlohner-Freedman bound. We are concerned with a Euclidean signature hyperbolic geometry, and so the dynamical instability associated to the Breitenlohner-Freedman bound is not relevant. As has been emphasised previously in \cite{deBoer:2003vf}, the relevant fact is rather that the principal series representation is unitary.

In addition to obeying the Laplace equation, the waveform is invariant under $SL(2,\Z)$.
This condition can be written in terms of the $SL(2,\R)$ generators in (\ref{eq:gen1}) as
\be\label{eq:ST}
e^{i H} \Psi_k = \Psi_k \qquad \text{and} \qquad e^{i\frac{\pi}{2}(H + K)} \Psi_k = \Psi_k \,.
\ee
Recall that $\frac{1}{2}(H+K) \equiv L_0$, which has a discrete spectrum.
The first of the conditions in (\ref{eq:ST}) is clearly invariance under the $T$ transformation $x \to x+1$. The second is invariance under the $S$ transformation $x + i y \to -1/(x+ i y)$. To see this, note from (\ref{eq:gen1}) that $i(H+K)$ generates a flow for $z(s) \equiv x(s) + i y(s)$ obeying
\be
\dot z = 1 + z^2 \,.
\ee
This differential equation has the solution
\be
z(s) = \frac{\sin(s) + z(0) \cos(s)}{\cos(s) - z(0) \sin(s)} \,,
\ee
which becomes the $S$ transformation at $s = \pi/2$.

We noted above that $SL(2,\Z)$ invariance together with oddness under $x \to -x$ implies vanishing of the waveform on the boundary of the half-fundamental domain. Therefore, in addition to the Laplace equation itself, the Dirichlet boundary conditions can also be imposed in an algebraic way. If we unfold the waveform to the entire upper half plane, using $SL(2,\Z)$ transformations and reflections about the $x=0$ axis, then we obtain a wavefunction on the $y=0$ conformal boundary. In the following subsections we explain how this boundary wavefunction can be characterised by imposing (\ref{eq:ST}) and oddness directly within a one dimensional conformal quantum mechanics. In \S\ref{sec:bdy} we show that the full bulk waveform can be reconstructed from the boundary wavefunction in a familiar way. The boundary description gives a less cluttered access to the key physical data, because it eschews the purely kinematic $K_k(\theta,t)$ functions of \S\ref{sec:autoL}.

Finally, we can define the Hecke operators using the $SL(2,\R)$ generators (\ref{eq:gen1}).
The Hecke relations (\ref{eq:hecke1}) and (\ref{eq:hecke2}), that express all Dirichlet coefficients in terms of the prime coefficients, follow from the existence of these operators. The prime Hecke operators $T_p$ are given by \cite{Bogomolny:1992cj}
\begin{equation}\label{eq:Tp}
    T_p \Psi_k(x,y)=\frac{1}{\sqrt{p}}\left(\Psi_k(px,py) + \sum_{j=0}^{p-1}\Psi_k\left(\frac{x+j}{p},\frac{y}{p}\right)\right) \,.
\end{equation}
In terms of the bulk $SL(2,\R)$ generators (\ref{eq:gen1}), we have
\be\label{eq:tp2}
T_p = \frac{1}{\sqrt{p}} \left(p^{- i D}  + p^{i D} \frac{e^{iH}-1}{e^{iH/p}-1} \right)
\,.
\ee
Here we used the simple identity $e^{i H}-1 = (e^{iH/p}-1) \sum_{j=0}^{p-1} e^{i j H/p}$.
The operators (\ref{eq:tp2}) manifestly commute with the Casimir, and therefore act within a representation. It can further be verified that the Hecke operators map automorphic forms to automorphic forms, and therefore preserve the modular invariant state (\ref{eq:ST}). The final term in (\ref{eq:tp2}) almost vanishes on invariant states, only picking up frequencies that are multiples of $p$
\be
\frac{e^{iH}-1}{e^{iH/p}-1} \sin(2 \pi n x) =
\left\{
\begin{array}{cr}
0 & p \nmid n\\
p \sin \left(\frac{2 \pi n x}{p} \right) & \text{else}
\end{array}
\right. \,.
\ee
This fact leads to the Hecke relations \cite{Bogomolny:1992cj}.

\subsection{Conformal quantum mechanics in the principal series}
\label{sec:cqm}

States in a principal series representation can be written in the `noncompact picture' \cite{knapp} as
\be
\ket{\psi} = \int dx \, \psi(x) \ket{x} \,,
\ee
where the wavefunction $\psi(x)$ has the usual $L^2$ norm. The $SL(2,\R)$ generators are realised as \cite{Anous:2020nxu}
\be\label{eq:hkd}
H\psi = - i \frac{d\psi}{dx} \,, \qquad D\psi = i \left(x \frac{d\psi}{dx} + \Delta \psi\right) \,, \qquad K\psi = - i \left(x^2 \frac{d\psi}{dx} + 2 x \Delta \psi\right) \,.
\ee
From these expressions one verifies that $\Delta$ indeed fixes the Casimir as in (\ref{eq:cas}). From (\ref{eq:dep}) we must then have $\Delta = \frac{1}{2} + i \vep$. For clarity we will drop the $k$ subscript on all quantities in our CQM discussion. It is understood, unless stated otherwise, that we are dealing with a specific waveform at a fixed value of $k$.

Given the generators (\ref{eq:hkd}), the $T$ and $S$ symmetries of the state in (\ref{eq:ST}) are seen to require
\be\label{eq:st2}
\psi(x) = \psi(x+1) \qquad \text{and} \qquad \psi(x) = \frac{1}{|x|^{2\Delta}} \psi\left(\frac{-1}{x}\right) \,. 
\ee
This is the transformation of a state in an {\it even} principal series representation \cite{knapp}. In obtaining the $S$ transformation in (\ref{eq:st2}) we restricted to odd parity wavefunctions obeying
\be\label{eq:xpx}
\psi(x) = - \psi(-x) \,,
\ee
corresponding to the odd automorphic waveforms (\ref{eq:odd}). 
Going forward, it will be simplest to use the oddness (\ref{eq:xpx}) to restrict attention to
\be
x \geq 0 \,,
\ee
together with the boundary condition $\psi(0)=0$. The generators (\ref{eq:hkd}) remain self-adjoint with these restrictions. Within this domain, $S$-invariance becomes
\be\label{eq:goodS}
\psi(x) = - \frac{1}{x^{2\Delta}} \psi\left(\frac{1}{x}\right) \,. 
\ee
This is the form we will use.

To implement the discrete translation invariance (\ref{eq:st2}) it will be easiest to work with non-normalisable `unfolded' states that are periodic in $x$.\footnote{A basis of {\it normalisable} states in an even principal series representation is given by the eigenfunctions of $L_0 = \frac{1}{2}(H+K)$ with eigenvalue $m \in \Z$. In our basis these eigenfunctions are
\be\label{eq:psim}
\psi_m(x) = \frac{e^{2 m i \arctan(x)}}{(1 + x^2)^\Delta} \,.
\ee
The quantisation of $m$ implies that $\lim_{x \to \infty} |x|^{2 \Delta} \psi_m(x) = \lim_{x \to -\infty} |x|^{2 \Delta} \psi_m(x)$. The eigenfunction is therefore well-defined on the `global' compact space obtained by setting $x = \tan\frac{\vartheta}{2}$ with $-\pi < \vartheta < \pi$, cf.~\cite{Spradlin:1999bn}.} We can think of these as scattering states. However, the vanishing of the wavefunction at $x=0$ together with the $T$ invariance in (\ref{eq:st2}) and the $S$ invariance in (\ref{eq:goodS}) implies that the wavefunction $\psi(x)$ vanishes at all rational numbers. It is therefore a highly irregular function. We will get an intuitive understanding of this fact in \S\ref{sec:bdy} below, where the wavefunction is obtained as the boundary value of the Maa{\ss} waveform. 

In the following \S\ref{sec:qm} we show that, in contrast, the wavefunction is smooth in a dilatation basis.
The eigenfunctions of the dilatation operator are, from (\ref{eq:hkd}) and $\Delta = \frac{1}{2} + i \vep$,
\be\label{eq:psit}
\psi_t(x) \equiv x^{- \frac{1}{2}- i (\vep+t)} \,, \qquad D \psi_t = t \, \psi_t \,.
\ee
With our restriction to $x \geq 0$, the inner product is
\be
\bra{\chi}\ket{\psi} = \int_0^\infty dx \, \overline{\chi}(x) \psi(x) \,.
\ee
The wavefunctions (\ref{eq:psit}) are recognised as conventional plane wave states upon making the coordinate change $x = e^{y}$ and accounting for the corresponding change in the measure.
Therefore, $t \in \R$ in (\ref{eq:psit}) gives a continuous spectrum of delta-function normalisable eigenfunctions.
The wavefunction in the basis of dilatation eigenstates is given by the Mellin transform
\be\label{eq:dil}
\phi(t) \equiv \int_0^\infty dx \psi(x) x^{- \frac{1}{2}+ i (\vep+t)} \,.
\ee
In this basis, invariance under the $S$ transformation (\ref{eq:goodS}) becomes
\be\label{eq:ref}
\phi(t) = - \phi(-t) \,.
\ee

\subsection{The wavefunction in the dilatation basis}
\label{sec:qm}

Requiring the wavefunction $\psi(x)$ to be invariant under the $T$ generator in (\ref{eq:ST}) and to be odd under $x \to -x$ means that we must have
\be\label{eq:an}
\psi(x) = \sum_{n=1}^\infty a_n \sin(2 \pi n x) \,.
\ee
Here the $a_n$ are, as yet, undetermined coefficients.
The dilatation basis wavefunction (\ref{eq:dil}) is then found to be
\be\label{eq:phit}
\phi(t)  = \Phi\left(\half + i t\right) \,,
\ee
where, using the reflection and duplication formulae for gamma functions,
\be\label{eq:Ps}
\Phi(s) = \frac{\sqrt{\pi}}{2\pi^{s+i\vep}}\frac{\Gamma\left(\frac{1+ s + i \vep}{2}\right)}{\Gamma\left(\frac{2 - s - i \vep}{2}\right)} \widetilde{L}\left(s\right) \,.
\ee
Here we set
\be
\widetilde{L}(s) \equiv \sum_{n=1}^\infty \frac{a_n n^{-i \vep}}{n^{s}} \,.
\ee

The reflection invariance (\ref{eq:ref}) is equivalent to $\Phi(s) = - \Phi(1-s)$. Using (\ref{eq:Ps}), this requires
\be\label{eq:Ltilde}
\frac{1}{\pi^{s}} \Gamma\left(\frac{1+ s + i \vep}{2}\right)\Gamma\left(\frac{1+s - i \vep}{2}\right) \widetilde{L}\left(s\right) = - \frac{1}{\pi^{1-s}}\Gamma\left(\frac{2-s + i \vep}{2}\right)\Gamma\left(\frac{2 - s - i \vep}{2}\right) \widetilde{L}\left(1-s\right)
\ee
Comparing with the definition of the xi function in (\ref{eq:xi}) and the reflection formula (\ref{eq:inv}), we see that this is precisely the condition obeyed by the automorphic $L$-functions.
There is a unique automorphic form at a given energy level $\vep$.
Thus, to solve the reflection condition we must set
\be\label{eq:ac}
a_n = n^{i\vep} c_n \,,
\ee
where $c_n$ are the Fourier coefficients of the Maa{\ss} waveform with eigenvalue $\vep$, so that
$\widetilde{L}(s) = L(s)$ and hence, using (\ref{eq:xi}),
\be\label{eq:phifin}
\phi(t) = \frac{\pi^{\frac{1}{2}-i \vep} \xi(\half + i t)}{\Gamma\left( \frac{3-2 i t- 2 i \vep}{4}\right) \Gamma\left( \frac{3+ 2 i t - 2i \vep}{4}\right)} \,.
\ee
It follows that the dilatation-basis wavefunction vanishes on precisely the zeros of the xi function:
\be\label{eq:connes}
\phi(t_n) = 0 \,.
\ee
The additional zeros of $\phi$ due to the inverse gamma functions in (\ref{eq:phifin}) are not at real $t$. As we have advertised above, the wavefunction (\ref{eq:phifin}) therefore loosely realises the suggestion by Connes that the zeroes of the $L$-function are an absorption spectrum. Furthermore, in the spirit of the Berry-Keating scenario, the spectrum in question is that of a dilatation operator. The zeroes are not, however, related to the density of states of the dilatation operator but rather the density of zeros. We will characterise the asymptotic behaviour of (\ref{eq:phifin}) in \S\ref{sec:semi} below.

\subsection{Interpretation as a boundary state}
\label{sec:bdy}

The wavefunction $\phi(t)$ can be obtained as the near-boundary behaviour of the unfolded automorphic waveform. The gamma functions in (\ref{eq:phifin}) arise directly from the near-boundary $\theta \to 0$ limit of the hypergeometric function in (\ref{eq:k}), dropping the $k$ labels,
\be
K(\theta,t) \approx \text{Re} \, \frac{\sqrt{\pi} \, \Gamma(-i\vep) \theta^{\frac{1}{2} + i \vep}}{\Gamma\left( \frac{3- 2 i t- 2 i \vep}{4}\right) \Gamma\left( \frac{3+ 2 i t - 2 i \vep}{4}\right)}  \,.
\ee
Therefore from (\ref{eq:inverse}) we obtain the CQM wavefunction (\ref{eq:phifin}) as the boundary value of the `bulk' waveform, Mellin transformed along the boundary direction,
\be\label{eq:bdymap}
\phi(t) = \frac{2 \pi^{- i \vep}}{\Gamma(- i \vep)} \int_0^\infty \lim_{\theta \to 0}\left[\Psi(\rho,\theta)\right]^\text{+} \rho^{-1-it} d \rho \,.
\ee
Here $[\cdots]^\text{+}$ instructs us to take the coefficient of the `plus' term, that behaves like $\theta^{\frac{1}{2} + i \vep}$ as $\theta \to 0$. There is some arbitrariness in the precise pre-factor in (\ref{eq:bdymap}) as neither the waveform nor the wavefunction have been normalised.

The near-boundary limit can also be taken in terms of the $x$ and $y$ coordinates. Expanding the waveform (\ref{eq:fourier}) at small $y$ gives
\be\label{eq:psilimit}
\psi(x) = \frac{2 \pi^{- i \vep}}{\Gamma(- i \vep)} \lim_{y \to 0}\left[\Psi(x,y)\right]^\text{+} \,,
\ee
where now $[\cdots]^\text{+}$ instructs us to take the coefficient of $y^{\frac{1}{2}+i \vep}$ near the boundary, and we used the relation (\ref{eq:ac}) between the $a_n$ and $c_n$ coefficients.

The operation in (\ref{eq:psilimit}) can be inverted to reconstruct the full waveform (\ref{eq:fourier}) in terms of the CQM wavefunction (\ref{eq:an}) as
\be\label{eq:ii}
\Psi(x,y) = \frac{\Gamma(\frac{1}{2} - i \vep)}{2 \pi^{\frac{1}{2} - i \vep}} \int_{-\infty}^\infty \frac{y^{\frac{1}{2} - i \vep} \psi(x') dx'}{\left[(x-x')^2 + y^2 \right]^{\frac{1}{2} - i \vep}} \,.
\ee
To recover (\ref{eq:psilimit}) from (\ref{eq:ii}), change variables to $x' = x + y x''$ and then take the $y \to 0$ limit. Similarly, in $(\rho,\theta)$ coordinates we can see that (\ref{eq:auto}) reconstructs the bulk waveform from the boundary dilatation basis wavefunction $\phi(t) \propto \xi(\frac{1}{2} + i t)$. Indeed, because the bulk dilatation operator in (\ref{eq:gen1}) is simply $D = i \rho \pa_\rho$ then we can directly interpret (\ref{eq:auto}) as giving the waveform in a basis of bulk dilatation eigenstates.

\begin{figure}[h]
    \centering
    \raisebox{-\height}{\includegraphics[height=0.55\linewidth]{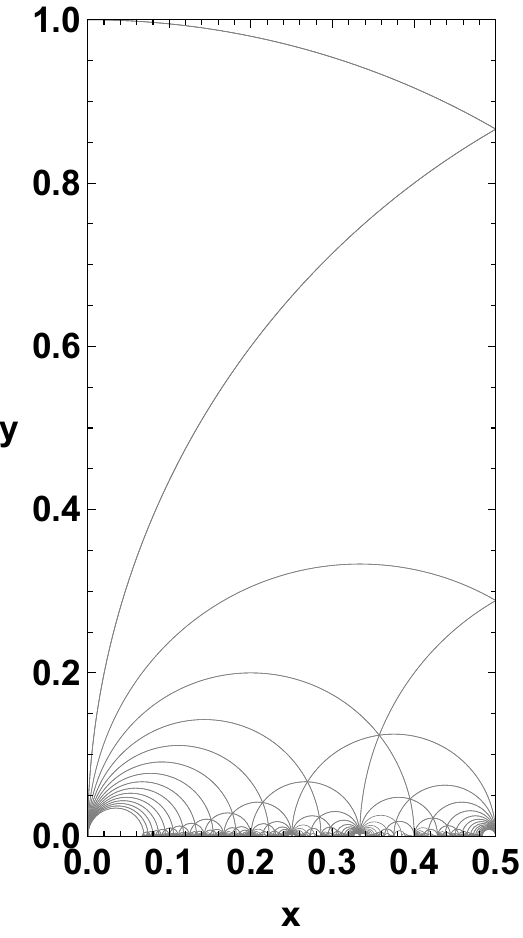}}
    \hspace{2cm}
    \raisebox{-\height}{\includegraphics[height=0.5\linewidth]{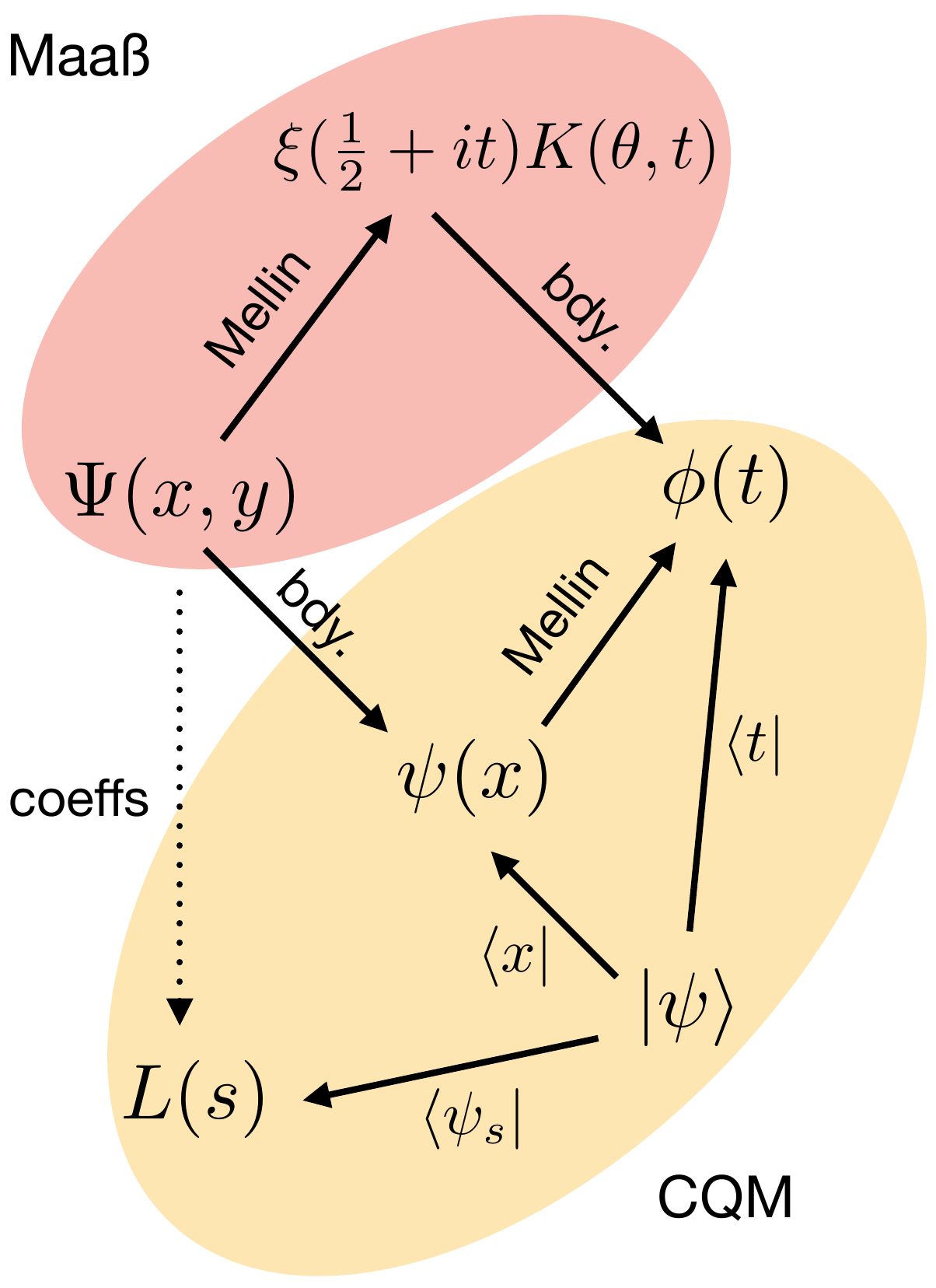}}
    \caption{{\bf Left:} The bulk waveform vanishes along $SL(2,\Z)$ domain boundaries, shown as lines in the figure. These accumulate towards the conformal boundary of the hyperbolic plane at $y = 0$, such that the boundary wavefunction $\psi(x)$ vanishes at all rational numbers. See the discussion in \cite{schmid2000automorphic,Miller2004}. {\bf Right:} relationship between various representations of the state. In the bulk, the Maa{\ss} waveform is first defined as $\Psi(x,y)$. Writing in bulk polar coordinates $(\rho, \theta)$ and Mellin transforming $\rho \leftrightarrow t$ gives the $\xi$ function along the critical axis.
    The boundary value of the Maa{\ss} waveform gives the wavefunction $\psi(x)$ of a state $\ket{\psi}$ in CQM in the position basis $|x\rangle$. The wavefunction $\phi(t)$ in a basis of dilatation eigenstates $\ket{t}$ is proportional to the $\xi$ function. A representation in terms of certain polylogarithmic wavefunctions $\ket{\psi_s}$ gives the $L$-function on the real axis. This $L$-function is built using the Fourier coeffcients of the original Maa{\ss} waveform.}
    \label{fig:modular}
\end{figure}

Viewing the CQM wavefunction as the boundary value of the bulk waveform, as we have just described, elucidates the irregular structure of $\psi(x)$ noted in \S\ref{sec:cqm} above. The odd Maa{\ss} waveform vanishes on $SL(2,\Z)$ domain boundaries. In Fig.~\ref{fig:modular} we recall how these boundaries accumulate as $y \to 0$. This leads to an accumulation of zeroes in $\psi(x)$. This figure also illustrates the relationship between the various bulk and boundary quantities that we have encountered.

\subsection{The $L$-function on the real axis as an overlap}
\label{sec:overlap}

In (\ref{eq:connes}) we have seen that the wavefunction in the dilataton basis is proportional to the $L$-function along the critical axis in the complex plane. Here we will show how $L(s)$ along the real axis can also be obtained from the wavefunction. This is of interest because for $s>1$ the $L$-function can be represented as the Euler product (\ref{eq:local}). In \S \ref{sec:ds} below we will see that this representation allows $L(s)$, and hence the wavefunction, to be interpreted naturally as a partition function of an auxiliary, or dual, primon gas.

The $L$-function along the real axis can be obtained from the overlap
\be\label{eq:Loverlap}
L(s) = \bra{\psi_s}\ket{\psi} \,,
\ee
where
\be\label{eq:phis}
\psi_s(x) \equiv 2 \, \text{Im} \, \text{Li}_s(e^{2 \pi i x}) \Theta(x) \Theta(1-x) \,.
\ee
The Heaviside step functions in (\ref{eq:phis}) means that the overlap (\ref{eq:Loverlap}) involves the integral over $x \in [0,1]$ only. The overlap can equivalently be thought of as involving only the normalisable `folded' wavefunction that is supported on $x \in [0,1]$.
The overlap is most easily computed by
writing the polylogarithm as
\be
\text{Im} \, \text{Li}_s(e^{2 \pi i x}) = \sum_{k=1}^\infty
 \frac{\sin (2 \pi k x)}{k^s} \,.
 \ee
The wavefunctions in (\ref{eq:phis}) are not orthogonal, with
\be
 \bra{\psi_s}\ket{\psi_{s'}} = 2 \zeta(s + s') \,.
\ee

\section{Relations between zeros and Fourier coefficients}
\label{sec:semi}

In the previous section we have seen how three natural ways of packaging the automorphic data correspond to different representations of a single CQM state, as illustrated in Fig.~\ref{fig:modular}. The position basis wavefunction $\psi(x)$ is a Fourier series built from the $\{c_n\}$. The dilatation basis wavefunction $\phi(t)$ is built using the zeros $\{t_n\}$ of the $L$-function along the critical line. In this section we use the position and dilatation space representations to relate the zeros and the Fourier coefficients.
In the following \S\ref{sec:ds} we will return to the Euler product representation of $L(s)$ along the real axis, which was obtained above from the overlaps of the wavefunction with
$\psi_s(x)$. While the `approximate functional equation' discussed in the following subsection relates the zeros to all of the $\{c_n\}$, the `explicit formula' discussed in \S\ref{sec:explicit} relates the zeroes directly to the prime coefficients and hence the $\{\theta_p\}$. We will drop the $k$ subscript throughout.

\subsection{Approximate functional equation}

This subsection will give an expression for the $L$-function along the critical line in terms of the $c_n$ coefficients of the original Dirichlet series. Such an expression is called
an `approximate functional equation', generalising the Riemann-Siegel formula for the Riemann zeta function to automorphic $L$-functions. An elegant discussion of these approximations for general $L$-functions can be found in \cite{conrey}. Our discussion will be heuristic, but we will corroborate the proposal numerically. The following \S\ref{sec:inver} will invert this result to obtain the Dirichlet series coefficients in terms of the critical line $L$-function, giving a strong check on the formula.

The first step is to isolate the phase of the $L$-function along the critical axis. We noted below (\ref{eq:zeros}) that the xi function is pure imaginary along the critical axis. Using the definition of the xi function in (\ref{eq:xi}), with $s = \frac{1}{2} + i t$ along the critical axis, we may therefore write
\be\label{eq:TZ}
L(\half + i t) \equiv e^{i \Theta(t)} Z(t) \,,
\ee
with $Z(t)$ real and the phase $\Theta(t)$ of the $L$-function along the critical axis being the odd function
\be
\Theta(t) = \frac{1}{2 i} \log \left[- \pi^{2 i t} \frac{\Gamma\left(\frac{3 - 2 i t - 2 i \vep}{4}\right) \Gamma\left(\frac{3 - 2 i t + 2 i \vep}{4}\right)}{\Gamma\left(\frac{3 + 2 i t + 2 i \vep}{4}\right) \Gamma\left(\frac{3 + 2 i t - 2 i \vep}{4}\right)} \right] \,.
\ee
The asymptotic behaviour of the phase at large positive $t$ is then found to be given by
\be\label{eq:ThetaAsymp}
\Theta(t) = - t \log \frac{t}{2 \pi e} + \frac{\pi}{4} + \cdots \,.
\ee
This asymptotic behaviour of the phase is twice that of the Riemann zeta function, reflecting the fact that the asymptotic zeros are twice as dense.

The Dirichlet series (\ref{eq:L}) does not converge along the critical line. One may hope to get an approximation to the correct answer by truncating the sum at some $N$th term. The essential point of the approximate functional equation is to improve this truncation by enforcing that the truncated series have, to the same approximation, the correct phase. Thus we write
\begin{align}
L(\half + i t) & = \sum_{n=1}^N \frac{c_n n ^{- i t}}{\sqrt{n}} + 
e^{2 i \Theta(t)}\sum_{n=1}^N \frac{c_n n ^{i t}}{\sqrt{n}} + \text{(error terms)} \\
& = e^{i \Theta(t)} \sum_{n=1}^N \frac{2 c_n}{\sqrt{n}} \cos \left(t \log n + \Theta(t) \right) + \text{(error terms)} \label{eq:RS}
\,.
\end{align}
The first term on the right is the na\"ive Dirichlet series on the critical line, and the second term is the improvement. For the case of the Riemann zeta function, the improvement term can be argued for by Poisson resummation of higher $n$ terms \cite{berryearly,berry1999riemann}. This suggests that $N$ should not be too large, as including both the original terms and their resummation would risk double counting. We now turn to this point.

\begin{figure}[h]
    \centering
    \includegraphics[width=0.65\linewidth]{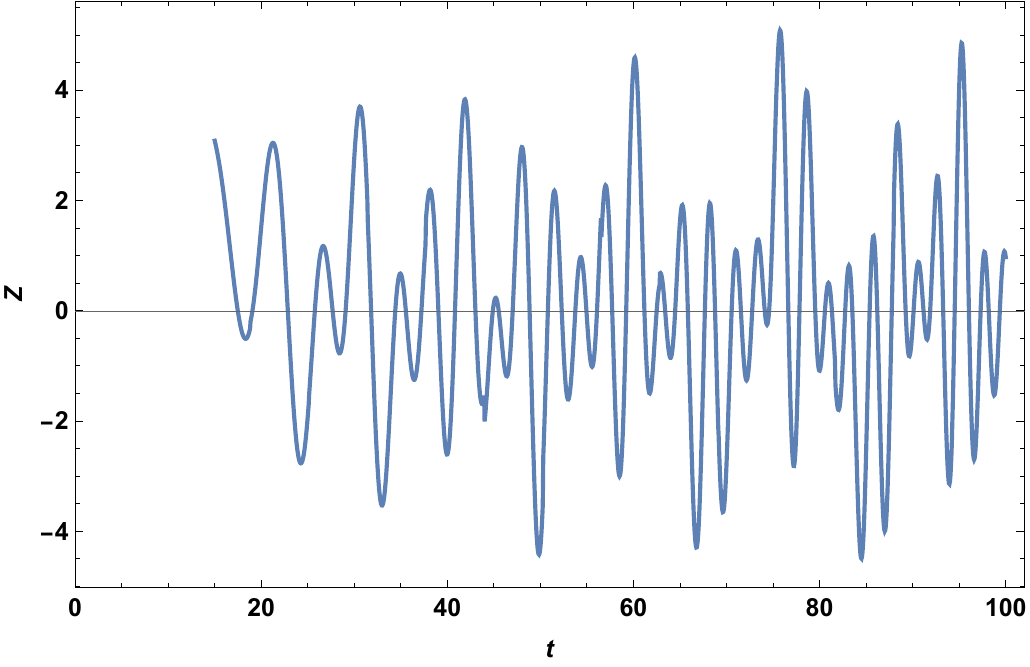}
    \caption{The function $Z(t)$ for the lowest energy odd Maa{\ss} waveform, with $\vep \approx 9.53$, computed using the approximate functional equation (\ref{eq:approx}) and numerical $c_n$ coefficients from the database \url{https://www.lmfdb.org/}. We have excluded low values of $t$ from the plot, as the approximation is not expected to be accurate there.}
    \label{fig:RS}
\end{figure}

Consider the approximation (\ref{eq:RS}) integrated over $t$ against some test function that is supported on $t < T$. The contribution to the integral from large $n$ terms in the sum (\ref{eq:RS}) can be obtained with a stationary phase approximation. Using the large $t$ expansion (\ref{eq:ThetaAsymp}) for the phase $\Theta(t)$, the stationary phase point for the $n$th term is found to be at
\be
t_{\star,n} = 2 \pi n \,.
\ee
An accurate computation of the integral should include all stationary phase points with $t_{\star,n} < T$. To capture these accurately we need to include terms in the sum up to $N = T/(2 \pi)$. This suggests the expression
\be
L(\half + i t) = e^{i \Theta(t)} \sum_{n \leq t/(2\pi)} \frac{2 c_n}{\sqrt{n}} \cos \left(t \log n + \Theta(t) \right) + \text{(error terms)} \label{eq:approx}
\,,
\ee
which is indeed the correct approximate functional equation for our class of $L$-functions \cite{conrey}. In the following \S\ref{sec:inver} we will see very explicitly the need for the cutoff $n \leq t/(2\pi)$ on the sum.
In Fig.~\ref{fig:RS} we have used (\ref{eq:approx}), with numerical Dirichlet coefficients
for the $L$-function corresponding to the lowest odd Maa{\ss} waveform taken from \url{https://www.lmfdb.org/}, to plot $Z(t)$.
From this plot we can read of all of the zeros with $t < 100$. This only requires the first fifteen Dirichlet coefficients. The zeros obtained in this way are found to agree with those given in \cite{booker}, obtained by a different method, to three significant figures (the first couple of zeros only agree to two significant figures). Given that the first thousand Dirichlet coefficients are known, the plot in Fig.~\ref{fig:RS} could easily be extended to find many more zeros.

\subsection{Formula for the Fourier coefficients}
\label{sec:inver}

In this subsection we obtain an expression for the Fourier coefficients of $\psi(x)$, which are also the Dirichlet coefficients of the $L$-function, as an integral of the $L$-function along the critical line. This can be thought of as an inverse of the approximate functional equation (\ref{eq:approx}).

An inverse Mellin transform of (\ref{eq:dil}) gives
\be\label{eq:melinv}
\psi(x)= \frac{1}{2 \pi} \int_{-\infty}^\infty dt \phi(t) x^{-\frac{1}{2} - i(\vep + t)} \,.
\ee
The $n$th Fourier coefficient can therefore be extracted as
\begin{align}
c_n = n^{-i \vep} a_n & = 2 n^{-i \vep}\int_0^1 dx \, \psi(x) \sin(2 \pi n x) \nonumber \\
& = \frac{n^{-i \vep}}{\pi} \int_{-\infty}^\infty dt \, \phi(t) F_n(\vep + t) = \frac{n^{-i \vep}}{\pi} \int_{0}^\infty dt \, \phi(t) \left[ F_n(\vep + t) - F_n(\vep - t) \right] \,, \label{eq:tint}
\end{align}
where we used (\ref{eq:ref}) in the last step and introduced
\be\label{eq:Fn}
F_n(z) \equiv \int_0^1 dx x^{-\frac{1}{2}}  e^{- i z \log x} \sin(2 \pi n x) = \frac{4 \pi n}{3 - 2 i z} {}_1F_{2}\left(\frac{3 - 2 i z}{4}; \frac{3}{2}, \frac{7 - 2 i z}{4};-n^2 \pi^2\right) \,.
\ee
The ${}_1F_{2}$ function (not to be confused with the more familiar ${}_2F_{1}$ hypergeometric function) can be evaluated efficiently numerically. Thus (\ref{eq:tint}) gives a practical formula to obtain the Fourier coefficients given $\phi(t)$ which, we recall from (\ref{eq:phifin}), is proportional to the $L$-function along the critical line. Note that $F_n(-z) = F_n(z)^*$.

It is instructive to recover the Fourier representation (\ref{eq:an}) for $\psi(x)$ from the approximate functional equation. The large $n$ Fourier terms are computed by the large $t$ contribution to the integral (\ref{eq:melinv}). We focus on these. At large positive $t$ we obtain from (\ref{eq:phit}) and (\ref{eq:Ps}) that
\be\label{eq:pz}
\phi(t) \approx \frac{i}{2} e^{i \vep \log \frac{t}{2 \pi}} Z(t) \,.
\ee
Here we used the definition of $Z(t)$ in (\ref{eq:TZ}) and the asymptotic behaviour (\ref{eq:ThetaAsymp}) of the phase. We now use the approximate functional equation (\ref{eq:approx}) for $Z(t)$ in (\ref{eq:pz}) and focus on the contribution of the $n$th term. Using the reflection property (\ref{eq:ref}) for $\phi(t)$, the contribution of this term to the integral (\ref{eq:melinv}) for $\psi(x)$ is then seen to be
\be
\left. \psi(x) \right|_n \approx \frac{c_n}{\sqrt{x n} \pi} \int_0^\infty e^{i \vep \log \frac{t}{2 \pi x}} \cos\left( t \log \frac{t}{2 \pi e n} - \frac{\pi}{4}\right) \sin \left(t \log x \right) dt \,.
\ee
This integral may now be performed by stationary phase. There are four stationary points. At large $n$ these are at
\be\label{eq:tstar}
t_\star \approx \frac{2 \pi n}{x} \pm \vep \qquad \text{and} \qquad
t_\star \approx 2 \pi n x \pm \vep \,.
\ee
The first two of these stationary points contribute
\be\label{eq:AA}
\left. \psi(x) \right|_n \approx - \frac{a_n}{x^{2 \Delta}} \sin \frac{2 n \pi}{x} \,,
\ee
recall that $a_n = c_n n^{i \vep}$ and $2 \Delta = 1 + 2 i \vep$, and the latter two contribute
\be\label{eq:BB}
\left. \psi(x) \right|_n \approx a_n \sin \left(2 n \pi x \right) \,.
\ee
This second expression (\ref{eq:BB}) is immediately recognised as the expected Fourier term (\ref{eq:an}). In fact, after summing over $n$, (\ref{eq:AA}) and (\ref{eq:BB}) are equal due to the $S$ invariance (\ref{eq:goodS}) of the wavefunction. Either (\ref{eq:AA}) or (\ref{eq:BB}) will do, and the crucial point is not to overcount by including both of them. This is precisely what is enforced by the restriction on the sum in 
the approximate functional equation (\ref{eq:approx}): for a given value of $x \neq 1$, and at large $n$, only one of the pairs of stationary points in (\ref{eq:tstar}) is in the allowed range $2 \pi n \leq t$. Hence, only one pair of stationary points contributes. The calculation just performed strongly validates the approximate functional equation (\ref{eq:approx}).

The discussion in the previous paragraph shows that an alternative way to write the approximate functional equation would be to relax the $t$-dependence of the cutoff on the sum and instead include an overall factor of $\frac{1}{2}$, to correct for double counting stationary points.

\section{The conformal primon gas}
\label{sec:ds}

\subsection{Primon gas of charged bosons}

We have seen in \S\ref{sec:overlap} how $L(s)$ along the real axis can be extracted from the CQM wavefunction. In this section we will explain how this object can be interpreted as a partition function. As we noted in the introduction, this duality between a wavefunction and partition function can perhaps be thought of in the spirit of the quantum Hall/CFT \cite{RevModPhys.89.025005} or dS/CFT \cite{Strominger:2001pn, Maldacena:2002vr} correspondences.

The defining expression for the $L$-function in (\ref{eq:L}) is suggestive of a thermal partition function at inverse temperature $s$. The coefficients in the Dirichlet series are not all positive. As we see shortly, this is due to an imaginary chemical potential. The fact that the partition function admits a product representation strongly suggests an interpretation in terms of non-interacting excitations with individual partition functions $L^{(p)}_k(s)$ in (\ref{eq:local}). This is the essential observation behind the primon gas interpretation of the Riemann zeta function \cite{julia1990statistical}. Here we give a generalisation of that construction to the automorphic $L$-function (\ref{eq:L}). We call the resulting system the automorphic or conformal primon gas.
See \cite{bakas} for earlier work on more general primon gases.

The local $L$-functions in (\ref{eq:local}) can be written, using the factorisation (\ref{eq:chiral}), as
\be\label{eq:part}
L^{(p)}_k(s) = \sum_{m,n=0}^\infty e^{- s (m+n) \log p + i \theta^k_p (m-n)} = \Tr e^{- s H^{(p)} + i \theta^k_p Q^{(p)}}\,.
\ee
In the final equality we have recognised the expression as the partition function of identical bosonic particles and anti-particles, where the Hamiltonian and charge operators
\be\label{eq:hp}
H^{(p)} = \omega_p \left(a_p^\dagger a_p + b_p^\dagger b_p  \right) \,, \qquad Q^{(p)} = \left(a_p^\dagger a_p - b_p^\dagger b_p  \right) \,,
\ee
with creation operators $a_p^\dagger$ and $b_p^\dagger$ and single-particle energy
\be
\omega_p = \log p \,.
\ee
The conformal primon gas is at inverse temperature $s$ and chemical potential $i \theta_p^k/s$.

We may obtain a path integral representation of the partition function by performing the sums
in (\ref{eq:part}), followed by standard manipulations:
\begin{align}
L^{(p)}_k(s) & = \frac{e^{s \, \omega_p} }{4 \sinh \left[\frac{1}{2} \left(s  \, \omega_p + i \theta^k_p \right)\right]\sinh \left[\frac{1}{2} \left(s  \, \omega_p - i \theta^k_p \right)\right]} \\
& = \frac{e^{s \, \omega_p}}{(s  \, \omega_p)^2 + (\theta^k_p)^2} \prod_{l\neq0} \frac{(2 \pi l)^2}{(2 \pi l + \theta^k_p)^2 + (s  \, \omega_p)^2} \\
& \propto e^{s \, \omega_p} \prod_l \int d\phi_l d\phi_l^* e^{- \phi_l^* \left[(2 \pi l + \theta^k_p)^2 + (s  \, \omega_p)^2 \right] \phi_l} \label{eq:phiphib} \\
& \propto \int {\mathcal D}\phi_p{\mathcal D}\phi_p^* \exp{- \int_0^1 d\tau \left(\left|\dot \phi_p{}(\tau) + i \theta_p^k \phi_p(\tau)\right|^2 + (s  \, \omega_p)^2 \left|\phi_p(\tau)\right|^2 - s  \, \omega_p\right)} \,. 
\end{align}
In the final two terms we are not keeping track of an overall (infinite) normalisation in the path integral. The complex field $\phi_p(\tau)$ has been introduced by the Fourier decomposition
\be
\phi_p(\tau) \equiv \sum_{l=-\infty}^\infty \phi_l e^{i 2 \pi l \tau} \,,
\ee
so that $\phi_p(0) = \phi_p(1)$. The full $L$-function is obtained from the product over primes (\ref{eq:local}):
\be\label{eq:fullpath}
L_k(s) = {\mathcal N}\int {\mathcal D}\phi{\mathcal D}\phi^*
\exp{- \int_0^1 d\tau \Lag_{k,s}[\phi,\phi^*]} \,,
\ee
where ${\mathcal N}$ is an overall normalisation and
\be\label{eq:Lks}
\Lag_{k,s}[\phi,\phi^*] = \sum_{p\in \P}\left(\left|\dot \phi_p + i \theta_p^k \phi_p\right|^2 + (s  \, \omega_p)^2 \left|\phi_p\right|^2 - s  \, \omega_p \right) \,.
\ee

\subsection{Averaging over Dirichlet coefficients}
\label{sec:ave}

The challenge with (\ref{eq:fullpath}) and (\ref{eq:Lks}) is to get a handle on the $\theta_p^k$. As we have noted above, these are conjectured to follow a Sato-Tate $\sin^2\theta$ distribution for a given $k$ \cite{Sarnak1987, n5}. It is difficult to make use of this deep fact because the expression for the partition function involves both $\theta_p^k$ and $p$ and the relation between these quantities is not known. We can, nonetheless, turn a different aspect of the randomness to our advantage. For each fixed $p$, the $c_p^k$ coefficients are also essentially random and have a known (and proven) distribution over $k$  \cite{Sarnak1987, n5}
\be\label{eq:measure}
\mu_p(x) \equiv \sum_k \delta(c_p^k - x) = \frac{(1+p) \sqrt{4- x^2}}{2 \pi [(\sqrt{p} + \frac{1}{\sqrt{p}})^2 - x^2]} \,.
\ee
This is the Kesten-McKay distribution.
It is a $p$-adic (cf.~\S\ref{sec:padic} below) Plancherel measure  and appears in several other contexts including random graph and random matrix theory. From the partition function point of view, (\ref{eq:measure}) is a measure on the space of primon gas theories, labelled by $k$. The distribution (\ref{eq:measure}) is for a fixed $p$ and does not capture the correlations between different prime coefficients. However, we may consider the logarithm of the partition function, which is a sum over $p$. Each term in the sum can then be independently averaged. We can expect this averaging to pick out universal aspects of the system. As noted in the introduction, this step is very much in analogy to recent works extracting universal aspects of CFT$_2$ dynamics by averaging over microscopic data that, as in our case, is subject to a modularity constraint \cite{Collier:2019weq, Belin:2020hea, Chandra:2022bqq}.

Using the measure (\ref{eq:measure}) we obtain:
\begin{align}
\Big\langle \log L_k(s) \Big\rangle_k & = \sum_{p \in \P} \left\langle \log L^{(p)}_k(s) \right\rangle_k \\ 
& = - \sum_{p \in \P} \int dx \mu_p(x) \log(1-x p^{-s}+p^{-2s}) \label{eq:sec} \\
& = \sum_{p \in \P} \frac{p-1}{2}\log \left( 1-p^{-(2s+1)} \right)\,. \label{eq:finals}
\end{align}
The final expression here is more tractable. We observe that the sign in front of the logarithm has changed in going from the second to third line above. The averaging, therefore, seems to reveal that the chemical potentials impart a fermionic nature to the system, as we now explain.

The averaged logarithm (\ref{eq:finals}) can be interpreted as the logarithm of the Witten index of a system of non-interacting, uncharged fermionic oscillators. That is, if we set
\be\label{eq:W1}
\Big\langle \log L_k(s) \Big\rangle_k \equiv \log W(s) \equiv \sum_{p \in \P} \frac{p-1}{2} \log W^{(p)}(s)  \,,
\ee
then 
\be\label{eq:Ws}
W^{(p)}(s) = 1 - p^{-(2s+1)}   =  \Tr \left( (-1)^F e^{ - (2s+1) \widetilde H^{(p)}}\right) \,.
\ee
This expression says that for every prime $p$ there is an unoccupied and an occupied fermionic state, with $F=0,1$ respectively. The occupied fermionic state has single-particle energy $\log p$, as the bosons did previously. That is, the fermionic Hamiltonian
\be
\widetilde H^{(p)} = \omega_p c_p^\dagger c_p \,,
\ee
with anticommuting creation operator $c_p^\dagger$. In (\ref{eq:Ws}) the inverse temperature is now $2s+1$.

The expression (\ref{eq:Ws}) is a well-known relation between the Witten index of a fermionic primon gas and the reciprocal of local factors of the Riemann zeta function \cite{Spector1990}.\footnote{The mathematical expression of this fact is the inversion formula
\be
\frac{1}{\zeta(s)} = \sum_{n=1}^\infty \frac{\mu(n)}{n^s} \,,
\ee
where the M\"obius function $\mu(n)$ is $(-1)^k$ if $n$ is the product of $k$ distinct primes and $0$ if $n$ contains a repeated prime. This function might be thought of as the mathematical discovery of fermions in 1832!
} An important difference in the full Witten index (\ref{eq:W1})
is that the $p$th fermion appears with degeneracy $\frac{p-1}{2}$.

\begin{figure}[h]
    \centering
    \includegraphics[width=0.65\linewidth]{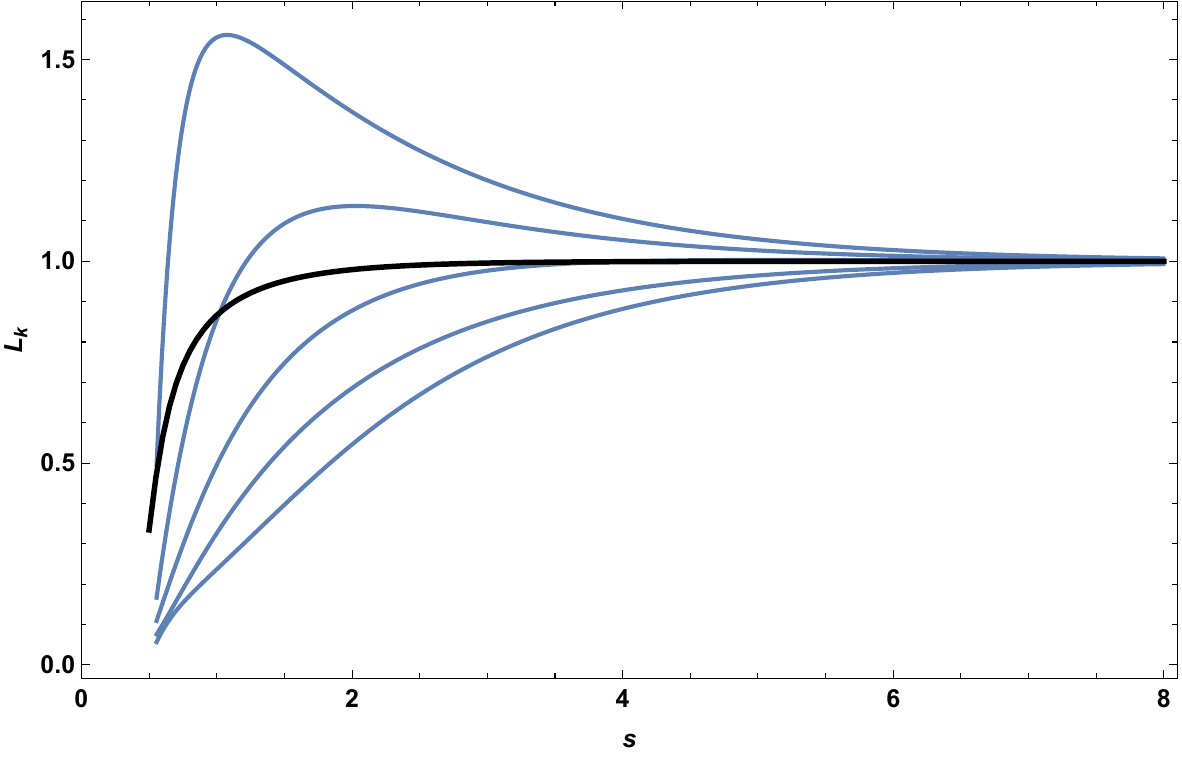}
    \caption{Thin blue lines: Several odd conformal primon gas partition functions $L_k(s)$, chosen to illustrate possible behaviours. From bottom to top: $\vep \approx \{110.17, 9.53,12.17,20.1,173.7 \}$. These plots use the first 1000 terms in the Dirichlet series, as given in \url{https://www.lmfdb.org/}. Thick black line: the average $e^{\langle \log L_k \rangle_k}$, plotted using (\ref{eq:finals}).}
    \label{fig:zplot}
\end{figure}

Several conformal primon gas partition functions, $L_k$, are plotted for $s \geq \frac{1}{2}$ in Fig.~\ref{fig:zplot}, together with the averaged quantity $e^{\langle \log L_k \rangle_k}$.
The individual partition functions vanish linearly at $s= \frac{1}{2}$. This is because odd automorphic $L$-functions must have a zero at $s = \frac{1}{2}$, as a consequence of the reflection symmetry (\ref{eq:inv}) of the xi function. Ordinarily, of course, partition functions must be positive. However, the conformal primon gas has imaginary chemical potentials. This zero in the partition function leads to a logarithmic divergence in $\log L_k \approx \log (s - \frac{1}{2})$. The averaged logarithm has a similar, but not identical, divergence as $s \to \frac{1}{2}$, also visible in Fig.~\ref{fig:zplot}. This divergence can be extracted from the large prime contribution to the sum (\ref{eq:finals}). Recall that the prime counting function $\pi(x) \sim \frac{x}{\log x}$. Expanding the logarithm in (\ref{eq:finals}) at large $p$, doing the integral and then taking $s \to \frac{1}{2}$, the large prime contribution is
\be\label{eq:diverge}
\sum_{p \gg 2} \frac{p-1}{2}\log \left( 1-p^{-(2s+1)} \right) \approx - \frac{1}{2} \int_{x \gg 2}^\infty dx \pi'(x) x^{-2s} \approx \frac{1}{2} \log \left(s - \frac{1}{2} \right) \,.
\ee
The additional factor of $\frac{1}{2}$ in front of the logarithm leads to a square root, rather than linear, vanishing of $e^{\langle \log L_k \rangle_k}$. This difference indicates that averaging does not commute with the limit $s \to \frac{1}{2}$. We believe that this occurs due to graphs like the top one in Fig.~\ref{fig:zplot}, with maxima that get increasingly close to the zero at $s=\frac{1}{2}$. The average of many such graphs can vanish more slowly as $s \to \half$ than any of the individual partition functions. We might rationalise this phenomenon as follows. Even and odd $L$-functions have the same statistical properties of Dirichlet coefficients, but
the even cases do not have a zero at $s=\frac{1}{2}$. The factor of $\frac{1}{2}$ in (\ref{eq:diverge}) is then due to an average of even and odd automorphic primon gases.

\begin{figure}[h]
    \centering
    \includegraphics[width=0.8\linewidth]{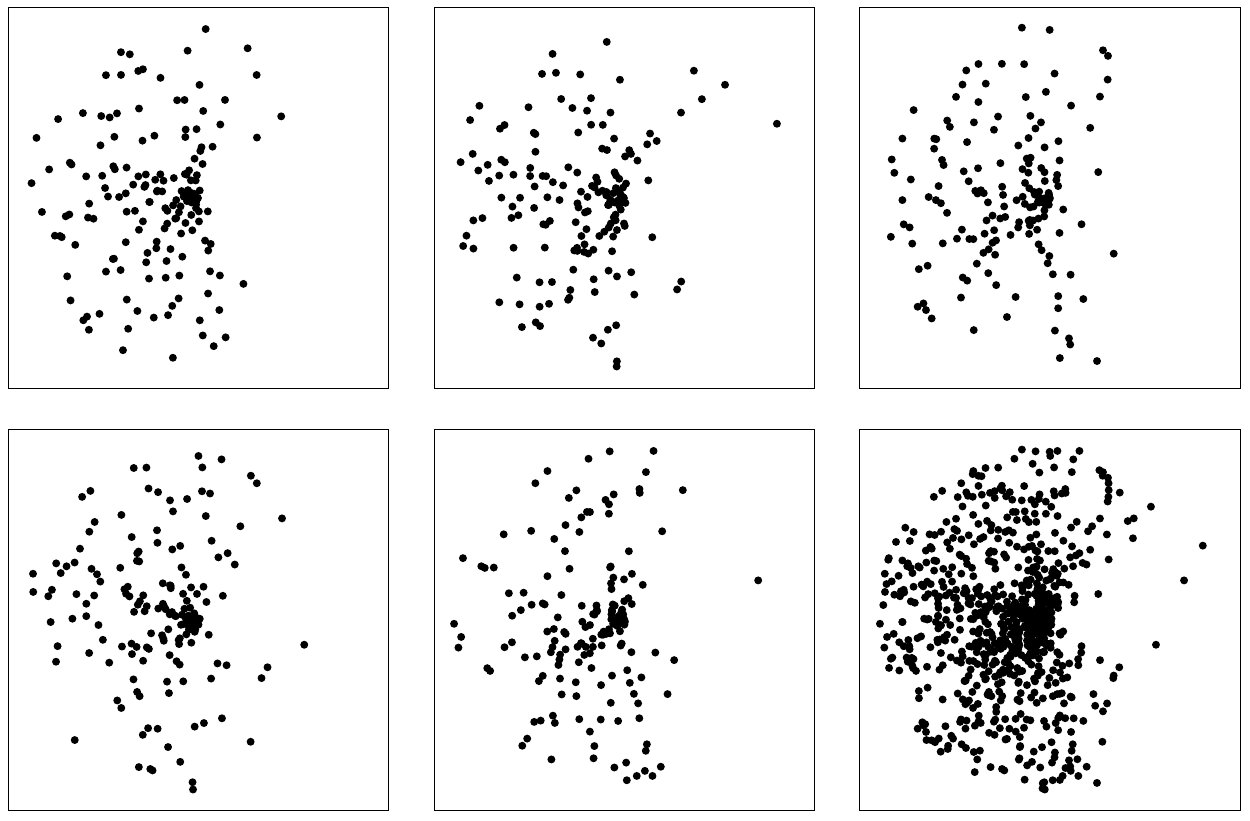}
\caption{The first five plots show the locations $\left(p \cos(2 \theta_p^k), p \sin(2\theta_p^k) \right)$ corresponding to the first 168 prime Dirichlet coefficients of the $L$-functions shown in Fig.~\ref{fig:zplot}. Starting on the top left and moving across, $\vep \approx \{9.53,12.17,20.1,110.17,173.7\}$. The final plot, bottom right, contains all of the points in the previous plots. As more individual $L$-functions are added, this plot will tend towards the distribution (\ref{eq:nu}) on each prime-radius circle.}
    \label{fig:circles}
\end{figure}

We end this subsection with a proposal for visualising the $L$-function data and their average. In terms of the phases $\theta^k_p$ the measure becomes
\be\label{eq:nu}
\nu_p(\theta) \equiv \sum_k \delta(\theta_p^k - \theta) = \frac{2}{\pi} \frac{p(1+p) \sin^2\theta}{1+p^2 - 2 p \cos(2 \theta)} \,.
\ee
Recall that $\theta \in [0,\pi]$. We can re-write this expression in terms of unit vectors $\vec x$ and $\vec y$:
\be
\nu_p(\theta) =   \frac{p(1 + p) (1 - \vec x \cdot \vec y)}{\pi \left(\vec x - p \, \vec y \, \right)^2} \quad \text{with} \quad
\left\{
\begin{array}{l}
\vec x = (1,0) \\ 
\vec y = \left(\cos(2 \theta), \sin(2\theta) \right) \,.
\end{array}
\right. \label{eq:xy}
\ee
The denominator in (\ref{eq:xy}) suggests associating a given $L_k$-function, which has a specific set of phases $\{\theta^k_p\}$, to the set of points on the plane $\{(p \cos(2 \theta^k_p),p \sin(2 \theta^k_p))\}$. Examples of such sets are shown in Fig.~\ref{fig:circles}. The angular distribution (\ref{eq:nu}) is then obtained, for each $p$, by combining the points of many distinct $L$-functions. This is also illustrated in Fig.~\ref{fig:circles}.

\subsection{Primon gas of charged fermions}

In the previous subsection we saw that suitably averaging over bosonic conformal primon gases produced the Witten index of a fermionic primon gas. In this subsection we explain how this correspondence also works in the other direction. We will start by showing that each individual $L$-function computes the Witten index of a primon gas of charged fermions. We will then see that averaging over these Witten indices produces the logarithm of the partition function of a bosonic primon gas.

Consider, again, non-interacting particles with Hamiltonian and charge given by (\ref{eq:hp}). Now, however, we take both the $a_p$ and $b_p$ oscillators to be fermionic. The inverse temperature and chemical potentials are taken as previously. By considering the possible occupied and unoccupied states, the Witten index of this theory is seen to be precisely the reciprocal of the $L$-function
\begin{align}
W_k(s) & \equiv \Tr \left( (-1)^F e^{\sum_p \left[ - s H^{(p)} + i \theta^k_p Q^{(p)} \right]} \right) \\
& = \prod_{p \in \P} \left(1 - e^{-s \log p + i \theta_p^k} - e^{-s \log p - i \theta_p^k} + e^{-2 s \log p} \right) = \frac{1}{L_k(s)} \,.
\end{align}

Averaging the logarithm of the Witten index proceeds identically to before except that, because we now have the reciprocal of the $L$-function, there is an additional minus sign in front of the final expression (\ref{eq:finals}). This allows the averaged index of the fermionic theory to be interpreted as a sensible bosonic partition function
\be
\left\langle \log W_k(s) \right\rangle_k \equiv \log Z(s) \,,
\ee
so that, from (\ref{eq:finals}),
\be\label{eq:Zs}
Z(s) = \prod_{p \in \P} \left[\frac{1}{1 - p^{-(2s+1)}}  \right]^{\frac{p-1}{2}} \,.
\ee
A plot of this function is the reciprocal of the averaged partition function shown in Fig.~\ref{fig:zplot}. In particular, it has a divergence rather than a zero as $s \to \frac{1}{2}$. This allows us to ascribe conventional thermodynamics to the averaged system.
As $s$ is the inverse temperature, the behaviour $\langle \log L_k(s)\rangle_k \approx - \frac{1}{2} \log \left(s - \frac{1}{2} \right)$ for the logarithm of the partition function --- note the opposite sign to (\ref{eq:diverge}) --- corresponds to a divergence of the energy as
\be\label{eq:en}
\left\langle E \right\rangle_k \approx \frac{1}{2} \frac{1}{s - \frac{1}{2}} \,.
\ee
From (\ref{eq:en}) we obtain a Hagedorn growth in the density of states at large $E$,
\be\label{eq:hag}
\rho(E) \propto e^{\frac{1}{2} E - \frac{1}{2} \log E} \,.
\ee

This expression for $Z(s)$ in (\ref{eq:Zs}) is somewhat similar to the uncharged primon gas partition function \cite{julia1990statistical}. As previously, the extra ingredient is the $\frac{p-1}{2}$ degeneracy of each oscillator. Repeating the steps leading to the path integral expression (\ref{eq:phiphib}) above, the averaged partition function is seen to describe the action
\be\label{eq:z2}
S = \sum_{p \in \P} \sum_{l=-\infty}^\infty \sum_{m=1}^{\frac{p-1}{2}} \phi_{l pm} \Big[(2 \pi l)^2 + (2s+1)^2 \omega_p ^2\Big] \phi_{lpm} \,.
\ee
Here the $\phi_{lpm}$ are now real and carry an extra index $m$ accounting for the degeneracy. The $p=2$ term should strictly be treated separately, as $\frac{2-1}{2} = \frac{1}{2}$ is not integer. We are shortly going to focus on the large $p$ limit. The degeneracy in $m$, being linear in $p$, is reminiscent of the azimuthal angular momentum quantum number, and suggests a two dimensional interpretation of the integral. We can make this manifest in the large $p$ limit. In this limit we can treat $p$ and $m$ as continuous variables and set $m = p p_\theta/(4 \pi)$, so that $p_\theta$ runs from 0 to $2\pi$. We may then introduce the two dimensional vector $\vec p = p (\cos p_\theta, \sin p_\theta)$. The action (\ref{eq:z2}) becomes
\be\label{eq:cont}
S_{p \gg 1} = \sum_{l = -\infty}^\infty \int \frac{d^2p}{4 \pi} \phi_{l\vec p} \left[\frac{(2 \pi l)^2}{\log |\vec p \,|} + (2s+1)^2 \log |\vec p \,| \right] \phi_{l \vec p} \,.
\ee
The additional inverse factor of $\log p$ comes from the leading order behavior of the asymptotic density of primes $\pi'(x) \sim 1/\log x$. The continuum 2+1 dimensional action (\ref{eq:cont}) captures the Hagedorn growth (\ref{eq:hag}) of the density of states.

\subsection{Averaged chemical potentials and the explicit formula}
\label{sec:explicit}

In this subsection we will describe a version of the explicit formula for automorphic $L$-functions. This formula relates the zeros of the $L$-function and the prime Dirichlet coefficients. This is different from \S\ref{sec:semi}, where the zeros were obtained
via the approximate functional equation (\ref{eq:approx}) which involves all of the Dirichlet coefficients. The prime coefficients are the basic data of the corresponding primon gas partition function.
By averaging over partition functions as in \S\ref{sec:ave} we obtain an (averaged) sum rule for the nontrivial zeros.

Using the Euler product representation (\ref{eq:chiral}) one obtains
\be\label{eq:dlogs}
\frac{d}{ds} \log L(s) = - \sum_{p \in \P} \sum_{n=1}^\infty \frac{2 \cos (n\theta_p) \log p}{p^{ns}} \,.
\ee
If $L(s)$ is thought of as a partition function, this derivative is the internal energy. The idea now is to perform an integral transform of (\ref{eq:dlogs}), such that the left hand side becomes an explicit function of the zeros, while the right hand side remains a sum over the chemical potentials. Taking the Bromwich integral we obtain
\be\label{eq:dlogs2}
\frac{L'(0)}{L(0)} + \sum_{s_\star} \frac{x^{s_\star}}{s_\star} = \frac{1}{2 \pi i} \int_{c - i \infty}^{c+i\infty} \left(\frac{d}{ds} \log L(s)\right) \frac{x^s}{s} ds = - \sum_{p^n \leq x} 2 \cos (n\theta_p) \log p \,.
\ee
Here we take $x > 1$. The sum on the left is over all of the zeroes $L(s_\star) = 0$, trivial and nontrivial. The first equality uses the fact that the derivative of the logarithm of $L$ is a sum of poles at $s=s_\star$, each with unit residue. In the middle expression, $c>1$ to be in the regime of absolute convergence of the Dirichlet series. The sum on the right is over all prime powers less than $x$. This final expression follows from using (\ref{eq:dlogs}) and considering the integral of each term in the sum separately. Depending whether $x/p^n$ is greater or less than one we may close the contour to the left or to the right, respectively. In the former case the integral picks up the residue from the pole at $s=0$.

From (\ref{eq:dlogs2}) one can obtain `explicit formulae' for automorphic $L$-functions.
In particular, differentiating with respect to $x$ gives a sum of delta functions on the right hand side
\be\label{eq:expl}
\sum_{s_\star} x^{s_\star - 1} = - \sum_{p^n} 2 \cos (n\theta_p) \log p \, \delta(x - p^n) \,.
\ee
This expression exhibits a Fourier-like relationship between the zeros and prime powers. The nontrivial zeros can be isolated by considering large $x$. Recall from
(\ref{eq:xi}) that the $L$-function has nontrivial zeros coming from the xi function, at $\frac{1}{2} + i t_n$, and trivial zeros coming from the gamma functions, at $-1 \pm i \vep - 2 \N$. Therefore,
\be\label{eq:nontrivial}
\sum_{s_\star} x^{s_\star - 1} = \frac{1}{\sqrt{x}} \sum_{n=-\infty}^\infty x^{i t_n}  + \sum_{m=1}^\infty \ocal\left(x^{-2-2m}\right) \,.
\ee
The first sum is over the nontrivial zeros at positive and negative $t_n$. At large $x$ the final contribution from the trivial zeroes is subleading. These terms are easily evaluated if needed.

The difficulty with using the expansion (\ref{eq:nontrivial}) in (\ref{eq:expl}) directly is that there are large fluctuations in the leading term as a function of $x$. These fluctuations can be smoothed out by averaging over the set of conformal primon gas partition functions, using the measure (\ref{eq:nu}).
The averaging is possible because the right hand side of (\ref{eq:expl})
is a sum over terms each involving a single prime $p$. It is easily seen that, for $n \geq 1$,
\be\label{eq:avv}
\int_0^\pi 2 \cos (n\theta_p) \nu_p(\theta) d\theta =
\left\{
\begin{array}{cl}
0 & \text{$n$ odd} \\[10pt]
\displaystyle \frac{1-p}{p^{n/2}} & \text{$n$ even}
\end{array}
\right. \,.
\ee
It can be verified by direct evaluation that performing the average (\ref{eq:avv}) on the right hand side of (\ref{eq:dlogs2}) produces a sum over prime powers that is a non-fluctuating function of $x$. This is to be contrasted with the highly fluctuating nature of, for example, the Chebyshev function $\psi(x) = \sum_{p^n \leq x} \log p$ arising in the corresponding expression for the Riemann zeta function. The smoother behaviour is due to the averaging over chemical potentials.

We therefore proceed to average (\ref{eq:expl}). Setting $n=2m$ we have
\be\label{eq:expl2}
\left\langle \sum_{s_\star} x^{s_\star - 1} \right\rangle_k = \sum_{p^{2m}} \frac{p-1}{p^m} \log p \, \delta(x - p^{2m}) \,.
\ee
The large $x$ limit is dominated by the $m=1$ term and can be evaluated to leading order as
\be\label{eq:nicesum}
\sum_{p^{2m}} \frac{p-1}{p^m} \log p \, \delta(x - p^{2m})
\approx \int dp \, \delta(x - p^2) = \frac{1}{2 \sqrt{x}} \,.
\ee
Here we used the asymptotic density of primes $\pi'(x) \approx \frac{1}{\log x}$. Putting (\ref{eq:nontrivial}), (\ref{eq:expl2}) and (\ref{eq:nicesum}) together we obtain
a sum rule for the averaged nontrivial zeros
\be
\lim_{x \to \infty} \left\langle \sum_{n=-\infty}^\infty  x^{i t_n} \right\rangle_k = \frac{1}{2} \,.
\ee
The averaging is crucial here, otherwise fluctuations render the limit undefined.

Retaining the trivial zeros, and thereby working at finite $x$, there is likely further information about the averaged nontrivial zeros that can be extracted from (\ref{eq:expl2}). We will leave that for future work.

\section{Future directions}

\subsection{Adelic perspective and $p$-adic holography}
\label{sec:padic}

The reflection formula (\ref{eq:inv}) for the xi function has an elegant interpretation via the adelic product. We will make a few brief comments about this perspective here.  The emergence of $p$-adic numbers in our setting, combined with a hyperbolic geometry, suggests that there may be connections to the notion of $p$-adic holography developed in \cite{Gubser:2016guj, Gubser:2017pyx}.

The adelic perspective on the $\xi$ function starts as follows. Each local factor $L^{(p)}_k(s)$ in the Euler product formula (\ref{eq:local}) is associated, in a way we describe below, to a `place' of $\mathbb{Q}$, with $p$-adic norm
\be
\left| \frac{a}{b} p^n \right|_p \equiv \frac{1}{p^n} \,, \label{eq:pnorm}
\ee
where the integers $a$ and $b$ contain no factors of $p$. The completion of $\mathbb{Q}$ under this norm is denoted $\mathbb{Q}_p$. The gamma function prefactor of the xi function in (\ref{eq:xi}) can then be associated to the infinite, or `Archimedean', place with the usual absolute value norm, now denoted $|\cdot|_\infty$, leading to the completion $\R$. 

As a basic example of an adelic product, the fundamental theorem of arithmetic may be written compactly as
\begin{equation}
    \prod_\nu |x|_\nu = 1 \,. \label{eq:product_norm}
\end{equation}
Here $\nu$ runs over the finite and infinite places. This expression is easily understood: the $p$-adic norms in the product cancel out the prime factors of $x$ appearing in $|x|_\infty$.
It is well-known that the reflection formula for the Riemann zeta function can be written succinctly in a form analogous to (\ref{eq:product_norm}), see e.g.~\cite{Huang:2020aao}. We now give an outline of how the xi function reflection formula (\ref{eq:inv}) can be written in a similar way.

The norms appearing in the product (\ref{eq:product_norm}) are instances of multiplicative characters, as they obey $|xy|_p = |x|_p |y|_p$. To write down the reflection formula (\ref{eq:inv}) we need more complicated multiplicative characters, given by
\be\label{eq:charac}
\chi_\infty(x) \equiv \text{sgn}(x) |x|_\infty^{s + i \vep} \,, \qquad \chi_p(x) \equiv |x|_p^s e^{i\theta_p \text{ord}_p(x)} \,.
\ee
Here $\vep$ and $\theta_p$ are the quantities we have encountered above and $\text{ord}_p(x)$ is the exponent $n$ appearing in (\ref{eq:pnorm}). That is, it counts the number of $p$ factors in $x$. We do not understand at this point how the expressions in (\ref{eq:charac}) should be combined into a `global' adelic character. Nonetheless, we will now see that that these characters encode the reflection formula.

We will also need the additive character $e(x) \equiv e^{2\pi i x}$, which can be generalised to $p$-adic fields \cite{Gubser:2016guj}. The additive and multiplicative characters can be combined to construct the local gamma factor
\be
    \Gamma_\nu \equiv \int_{\mathbb{Q}_\nu} \frac{d\mu(x)}{|x|_\nu}\, \chi_\nu(x) e_\nu(x) \,. \label{eq:gamma_factor}
\ee
We will not define the $p$-adic integral here -- see e.g.~\cite{Gubser:2016guj}. The $p$-adic integrals are perfomed by splitting the domain of integration into regions of a fixed $\text{ord}_p(x)$. The integrals evaluate to, using the duplication and reflection formulae of gamma functions for $\Gamma_\infty$,
\begin{align}
    \Gamma_p = \frac{1-e^{-i\theta_p}p^{s-1}}{1-e^{i\theta_p}p^{-s}} \,, \qquad
    \Gamma_\infty  = i\frac{\pi^{(1-s-i\vep)/2}\Gamma(\frac{s+i\vep+1}{2})}{\pi^{(s+i\vep)/2}\Gamma(\frac{2-s-i\vep}{2})} \,.
\end{align}
The reflection formula $\xi(s) = -\xi(1-s)$, written in the form (\ref{eq:Ltilde}) and with the $L$-functions written as (\ref{eq:chiral}), is then seen to be equivalent to the adelic product
\begin{equation}
    \prod_\nu \Gamma_\nu \bar{\Gamma}_\nu = 1 \,.
\end{equation}
Here the overline is not quite complex conjugation, but means send $\vep \to - \vep$ and $\theta_p \to - \theta_p$.

\subsection{Period function}

We have seen above how the CQM state $|\psi\rangle$ can be expressed in different bases. The complementary insights offered by the different representations is related to a certain tension between the $T$ and $S$ invariance of the wavefunction. The $T$ invariance is easily implemented in the position basis, as the periodicity of $\psi(x)$ in (\ref{eq:an}). The $S$ invariance is instead simplest in the dilatation basis, where it acts as the reflection (\ref{eq:ref}) of $\phi(t)$. In this section we will briefly introduce the so-called period function, which unifies the two invariances into a single 3-term functional equation \cite{period}. It seems likely that this function will be important for future studies of our CQM.

The period function is defined as
\begin{equation}
    \varphi(z) \equiv \int_0^\infty \frac{\psi(x) \, dx}{(z + x)^{2\bar{\Delta}}} \,, \label{eq:lewis_3pt}
\end{equation}
where $\bar \Delta = 1 - \Delta$. Using the $S$ and $T$ invariance of $\psi(x)$, as given in (\ref{eq:st2}) and (\ref{eq:goodS}) above, it is straightforward to show that the period function obeys the 3-term relation
\begin{equation}
    \varphi(z) = \varphi(z+1) - \frac{1}{z^{2\bar\Delta}} \varphi\left(\frac{z+1}{z}\right) \,. \label{eq:3term}
\end{equation}
This period function is furthermore seen to be holomorphic in \( \mathbb{C} \setminus (-\infty, 0] \) and obeys the growth conditions on the real axis that \( \varphi(x) = o(1) \) as \( x \to \infty \) and \( \varphi(x) = o(1/x) \) as \( x \to 0 \). It is shown in \cite{period} that, conversely, a unique odd Maa{\ss} waveform can be constructed from every solution to (\ref{eq:3term}) that is real analytic on $\R_+$ and obeys these growth conditions. The 3-term relation (\ref{eq:3term}), together with the growth conditions, therefore provides a succinct characterisation of the invariant CQM states.

\subsection{Superimposing waveforms and many-body CQM}

We have focused throughout on the physics of a single automorphic waveform. A dual description of BKL dynamics --- even at the semiclassical level --- will involve two further steps. Firstly, the states for distinct levels $\vep$ should be combined into a single theory. Secondly, recall that the automorphic billiard description arises independently at each point in the spatial slice undergoing BKL dynamics. These distinct points should also be combined into a single theory. We discuss these issues in turn.

Within the CQM of \S\ref{sec:conf} it is straightforward to amalgamate the states with differing $\vep$. Each state transforms in a distinct principal series representation, with the allowed representations fixed by the modular invariance conditions (\ref{eq:ST}). With all of the states at hand, one can form wavepackets that correspond to classical BKL spacetimes. It is possible that an uplift of the semiclassical dynamics to a finite quantum theory of the singularity should truncate the allowed `energies' $\vep$. Clearly, a more complete theory becomes necessary as the singularity is approached and the curvatures become large. It is natural to ask if the full theory preserves the conformal and automorphic structure of the semiclassical limit.

Each point in the spatial slice has its own set of automorphic states. The full theory must tensor all of these states together, leading to a many-body CQM. There are a couple of comments to make here. Firstly, from a fundamental point of view, Planck scale graininess and stringy effects should ultimately discretise the number of independent `points'. This may connect to the ideas developed in \cite{Chakravarty:2020wdm, Balasubramanian:2022gmo}. Secondly, while the distinct points/tensor factors are non-interacting deep within the BKL regime, at earlier times (away from the singularity) they will start to interact. These interactions are governed by the Einstein equation. This picture suggests that Einstein's equation emerges as a renormalisation group flow away from the CQM.
Unlike in holographic dictionaries anchored at a boundary, there is no UV/IR correspondence here. The non-interacting many-body CQM pertains in the regions of strong curvature and would therefore directly be the UV fixed point of the bulk dynamics.

Further to this last point, the hyperbolic nature of the gravitational superspace survives away from the strict BKL hard wall limit (\ref{eq:Ham1}). At earlier times the hard billiard walls are softened to time-dependent exponential potentials \cite{Damour:2002et, belinski_henneaux_2017}. The dynamics in this regime may be amenable to a CQM perspective.

The conformal primon gas picture of \S\ref{sec:ds} also admits a many-body generalisation in which distinct primon gas partition functions are tensored together. The resulting free energy (log of the partition function) is therefore a sum over the free enegy at each point in the spatial slice. It is possible that such an averaging over space leads, for generic states, to an averaging over energy levels similar to that considered in \S\ref{sec:ave} and \S\ref{sec:explicit}. On the other hand, superimposing states to form wavepackets does not seem natural from the point of view of a dual partition function --- similarly to the case of the dS/CFT correspondence.

\subsection{Superspace holography and conformal field theory}

The conformal boundary of asymptotically AdS spacetimes has underpinned progress in holography for decades \cite{Maldacena:1997re}. A non-gravitating clock can be placed at the boundary, endowing the gravitational system with a non-vanishing Hamiltonian and preferred notion of time. Cosmologies and black hole interiors do not have such a rigid boundary and need not have a preferred time. However, a choice of 3+1 decomposition can be made and the corresponding
Hamiltonian constraint can, in principle, be solved to express one of the field momenta in terms of the others. We may think of the field momentum picked out in this way as a relational Hamiltonian, generating translations in a relational time. The relational time could be, for example, the spatial volume or the conjugate York time. The DeWitt norm is preserved by the evolution generated by such a Hamiltonian, for any choice of relational clock. See also \cite{Witten:2022xxp} for a recent discussion of norms in this context.

We have reviewed above how, in the near-singularity BKL regime, 
there is a relational Hamiltonian (\ref{eq:billiard}) that is given by motion in hyperbolic space. For this choice of clock, then, the relational dynamics unfolds in a (Euclidean signature) AdS space. With a similar flavour to the AdS/CFT correspondence, we are suggesting that there is a dual description of this relational dynamics in terms of a many-body conformal quantum mechanics. We could call this `superspace holography'.

A major task for the future is to formulate a relational Hamiltonian that is dual not only to the semiclassical gravitational states, but also the full Hilbert space of quantum string theory close to a singularity. Our hope is that the powerful symmetries emerging in this regime --- conformal and modular invariance --- can be a guide in this endeavor. It is likely that higher dimensional cosmological billiards is the correct setting for a complete microscopic description
\cite{Damour:2002et,Kleinschmidt:2009hv, Kleinschmidt:2009cv,belinski_henneaux_2017,Kleinschmidt:2022qwl}.

One promising aspect are the similarities with the physics of two dimensional conformal field theories. We will now describe various senses in which the individual automorphic waveforms can be thought of as simplified systems with CFT$_2$-like structure. The highly developed understanding of CFT$_2$s may therefore be helpful in assembling and deforming these waveforms into a more complete theory. Firstly, recent work has used Maa{\ss} cusp forms as a natural modular-invariant basis for CFT$_2$ partition functions \cite{Benjamin:2021ygh, Haehl:2023tkr, DiUbaldo:2023qli, Haehl:2023wmr}. Relatedly, three dimensional quantum cosmology on a spatial torus also leads directly to these waveforms \cite{Godet:2024ich}. Secondly, we have seen that there are many solutions to the modular invariance conditions in CQM. While each solution appears largely random, the set of solutions follow nontrivial distributions. As we have noted, this is reminiscent of the emergence of structured randomness in CFT$_2$s \cite{Collier:2019weq, Belin:2020hea, Chandra:2022bqq}. Finally, there are multiple tantalising connections to random matrix theory. 
Both the set of prime phases $\{\theta_p\}$ and the set of nontrivial zeros $\{t_n\}$, for each state individually, have statistical properties related to random matrices. We have seen that these descriptions correspond to writing the same state in different bases. Understanding these quantities as the spectrum of a complicated operator in a CQM or CFT$_2$ would, of course, have far-reaching mathematical implications.

\section*{Acknowledgements}

SAH would like to thank the participants of the AfterStrings meeting at TIFR for comments on a talk on this material, and also Darius Shi and Steve Shenker for discussions, some years ago now, about the zeros of the Riemann zeta function. MY acknowledges helpful discussions with Jack Thorne. We also acknowledge helpful comments from Jon Keating, Axel Kleinschmidt and Tom Hartman on a first version.
This work has been partially supported by STFC consolidated grant ST/T000694/1. SAH is partially supported by Simons Investigator award $\#$620869. MY was supported by a Gates Scholarship ($\#$OPP1144).

\providecommand{\href}[2]{#2}\begingroup\raggedright\endgroup

\end{document}